\documentclass[aps,pre,twocolumn,superscriptaddress]{revtex4-1}

\usepackage[utf8]{inputenc}
\usepackage[T1]{fontenc}
\usepackage{lmodern}
\usepackage{amsmath,amsfonts,amssymb}
\usepackage{graphicx}

\usepackage{hyperref}
\usepackage[usenames,dvipsnames]{xcolor}
\hypersetup{colorlinks=true, linkcolor=blue!50!black, urlcolor=blue!50!black, citecolor=blue!50!black}
\usepackage[all]{hypcap}

\newcommand{\img}{\mathsf{i}}
\newcommand\diff{\mathrm{d}}
\newcommand{\Ps}{\text{Ps}}
\newcommand{\subref}[2]{\hyperref[#1]{#2}}
\renewcommand{\vec}[1]{\mathbf{#1}}
\renewcommand{\phi}[0]{\varphi}

\usepackage[normalem]{ulem}

\begin{document}

\title{Dynamically Crowded Solutions of Infinitely Thin Brownian Needles}


\author{Sebastian Leitmann}
\affiliation{Institut f\"ur Theoretische Physik, Universit\"at Innsbruck, Technikerstra{\ss}e~21A, A-6020 Innsbruck, Austria}
\author{Felix H\"ofling}
\affiliation{Fachbereich Mathematik und Informatik, Freie Universität Berlin, Arnimallee 6, 14195 Berlin, Germany}
\author{Thomas Franosch}
\affiliation{Institut f\"ur Theoretische Physik, Universit\"at Innsbruck, Technikerstra{\ss}e~21A, A-6020 Innsbruck, Austria}
\email[]{thomas.franosch@uibk.ac.at}

\date{\today}

\begin{abstract}
We study the dynamics of solutions of infinitely thin needles up to densities deep in the semidilute regime by Brownian
dynamics simulations. For high densities, these solutions become strongly entangled and the motion of a needle is
essentially restricted to a one-dimensional sliding in a confining tube composed of neighboring needles. From the
density-dependent behavior of the orientational and translational diffusion, we extract the long-time transport
coefficients and the geometry of the confining tube. The sliding motion within the tube becomes visible in the
non-Gaussian parameter of the translational motion as an extended plateau at intermediate times and in the intermediate
scattering function as an algebraic decay. This transient dynamic arrest is also corroborated by the local exponent of
the mean-square displacements perpendicular to the needle axis. Moreover, the probability distribution of the
displacements perpendicular to the needle becomes strongly non-Gaussian, rather it displays an exponential distribution
for large displacements. On the other hand, based on the analysis of higher-order correlations of the orientation we
find that the rotational motion becomes diffusive again for strong confinement. At coarse-grained time and length
scales, the spatiotemporal dynamics of the needle for the high entanglement is captured by a single freely diffusing
phantom needle with long-time transport coefficients obtained from the needle in solution. The time-dependent dynamics of
the phantom needle is also assessed analytically in terms of spheroidal wave functions. The dynamic behavior of the
needle in solution is found to be identical to needle Lorentz systems, where a tracer needle explores a quenched
disordered array of other needles. 
\end{abstract}

\pacs{87.15.hj, 87.15.H-, 66.10.C-}

\maketitle




\section{Introduction}

Solutions of rod-shaped particles such as filamentous actin
(f-actin)~\cite{Hinner:PRL_81:1998,Wong:PRL_92:2004,Liu:PRL_96:2006,Koenderink:PRL_96:2006},
microtubules~\cite{Lin:MA_40:2007}, xanthan~\cite{Koenderink:PRE_69:2004},
filamentous bacteriophage fd~\cite{Lettinga:PRL_99:2007, Grelet:PRX_4:2014}, and carbon nanotubes~\cite{Cassagnau:Rheol_52:2013} 
exhibit rich structural and dynamic
behavior~\cite{Bausch:NatPhys_2:2006,Solomon:SoftM_6:2010}.  Already at the level of a single constituent, the diffusive
motion of such an anisotropic particle is much more complex than a spherical one. While the long axis
undergoes rotational diffusion, translational diffusion is characterized by a parallel and a slower perpendicular
component with respect to the current orientation~\cite{Han:Science_314:2006, Duggal:PRL_96:2006,
Mukhija:JCIS_314:2007}. In solution, rod-shaped particles exhibit different concentration regimes depending on their
length $L$ and their diameter $b$~\cite{Doi:Oxford:1999}. In the dilute regime where the number density $n$ of particles is
very small, $nL^3 \ll 1$, the behavior of a single particle is not affected by its neighbors. The semidilute regime,
$nL^3 \gtrsim 1$, is characterized by a dynamic response due to entanglement effects of the particles with each other and
persists as long as the excluded volume of the individual particles is irrelevant, $n \ll 1/bL^2$.

For large aspect ratios $L/b$, the rods can be approximated by infinitely thin needles of length $L$ in the semidilute regime.  A remarkable
property of such solutions is their trivial ideal-gas-like static structure in striking contrast to their rich dynamic
behavior, since no two needles can cross each other. Deep in the semidilute regime, these solutions become dynamically
crowded and the dynamics of a single needle is restricted to a sliding motion within a tube formed by its neighbors. As
a consequence of this high entanglement, the rotational motion and the translational diffusion perpendicular to the needle
slow down drastically, whereas the diffusion along the tube is unaffected. In both
theory~\cite{{Doi:JP_36:1975},Teraoka:PRL_55:1985,Teraoka:JCP_89:1988,Teraoka:JCP_91:1989,Szamel:PRL_70:1993,Doi:Oxford:1999}
and computer simulations~\cite{Doi:JPSJ_53:1984,Cobb:JCP_123:2005,Tao:JCP_124:2006,
Tse:JCP_139:2013,Zhao:Poly_54:2013,Leitmann:PRL_117:2016} the density-dependent scaling behavior of the long-time
rotational and perpendicular translational diffusion coefficients have been established and scale with the number density as
$n^{-2}$. Computer simulations for two-dimensional toy models have also been performed
earlier~\cite{Moreno:EPL_67:2004,Hoefling:PRL_101:2008,Hoefling:PRE_77:2008,Munk:EPL_85:2009,Tucker:JPCA_114:2010} and
for a needle in the presence of pointlike obstacles one observes the same scaling laws of the transport
coefficients as in three dimensions~\cite{Hoefling:PRE_77:2008,Munk:EPL_85:2009}. In experiments, the transport coefficients
of a nanowire diffusing through an array of obstacles have been determined only recently, and the drastic slowing down of transport
has been observed~\cite{Kasimov:PRE_93:2016}.

The seminal tube concept for stiff fibers pioneered by Doi and Edwards~\cite{Doi:JCSFT_74:1978} furthermore reduces the
complex many-body dynamics of such solutions on coarse-grained time and length scales to that of a single needle
(phantom needle) with very unusual diffusion coefficients. We have shown recently~\cite{Leitmann:PRL_117:2016} that this
striking simplification is valid for the translation-rotation coupling as well as the intermediate scattering function
of the geometric center of the needle.

Here, we extend our earlier analysis~\cite{Leitmann:PRL_117:2016} and consider additional quantities characterizing the
dynamics of the needle in solution such as higher-order orientational correlation functions, mean-square displacements,
the non-Gaussian parameter of the geometric center, and the intermediate scattering function for a needle where each segment
contributes to the scattering signal. We also provide analytic formulas for the mentioned quantities in terms of the
phantom needle. In particular, for the non-Gaussian parameter in the highly entangled regime, we observe an extended
plateau over many decades in time which emerges due to the sliding motion of the needle within the confining tube. 

\section{Stochastic dynamics}

\subsection{Single needle}

We describe the configuration of a needle by its geometric center $\vec{r}$ and its unit vector of orientation,
$\vec{u}$.  The change in position, $\diff\vec{r}$, and orientation, $\diff\vec{u}$, of the needle is determined by the
following (overdamped) Langevin equations in It\={o} interpretation~\cite{Chirikjian:Birkhaeuser:2009,Oksendal:Springer:2010}:
\begin{align}   
\begin{split} \label{EqLangevinEquations}
  \diff\vec{u} &= -2 D_\text{rot}^0 \vec{u} \diff t - \sqrt{2 D_\text{rot}^0} \vec{u} \times \boldsymbol\xi \diff t, \\
  \diff\vec{r} &= \big[ \sqrt{2 D_\parallel^0} \vec{uu} + \sqrt{2 D_\perp^0} (1 - \vec{uu})
\big]\boldsymbol\eta \diff t ,
\end{split}
\end{align}
with rotational diffusion coefficient $D_\text{rot}^0$ and translational diffusion coefficients for parallel and
perpendicular motion, $D_\parallel^0$ and $D_\perp^0$, respectively. The dyadic product $\vec{uu}$ acts as a projector
onto the long axis $\vec{u}$ of the needle and introduces the coupling of translation and rotation. The stochastic
nature of the motion is modeled by the independent Gaussian white-noise processes $\boldsymbol\xi$ and $\boldsymbol\eta$
with zero mean and covariance $\langle \xi_i(t)\xi_j(t')\rangle = \langle \eta_i(t)\eta_j(t')\rangle = \delta_{ij}\delta(t-t')$.  

In computer simulations, we use a discrete fixed Brownian time step $\tau_\text{B}$ and implement the Langevin
equations~[Eq.~\eqref{EqLangevinEquations}] by evolving the needle ballistically in the time interval $\Delta t \in
[0,\tau_\text{B}]$ via the propagation rules: 
\begin{align}
\begin{split} \label{EqPropagationRules}
  \vec{u}(t + \Delta t) &= \vec{u}(t)\cos(|\boldsymbol\omega| \Delta t) +
\biggl(\frac{\boldsymbol\omega}{|\boldsymbol\omega|}\times \vec{u}(t)\biggr)\sin(|\boldsymbol\omega|
\Delta t),\\
  \vec{r}(t + \Delta t) &= \vec{r}(t) + \vec{v}\Delta t.
\end{split}
\end{align}
The random pseudovelocities $\boldsymbol\omega$ and $\vec{v}$ for rotational and translational motion,
respectively, are determined at the beginning of every Brownian step according to 
\begin{align}
\begin{split} \label{EqPseudoVelocities}
\boldsymbol\omega &= \sqrt{\frac{2 D_\text{rot}^0}{\tau_\text{B}}}(1 - \vec{uu})\boldsymbol{\mathcal{N}}_\xi ,\\
\vec{v} &= \Big[ \sqrt{\frac{2D_\parallel^0}{\tau_\text{B}}} \vec{uu} + \sqrt{\frac{2D_\perp^0}{\tau_\text{B}}} (1 -
\vec{uu}) \Big]\boldsymbol{\mathcal{N}}_\eta, 
\end{split}
\end{align}
where the random variables $\boldsymbol{\mathcal{N}}_\xi$ and $\boldsymbol{\mathcal{N}}_\eta$ are drawn from a normal
distribution with zero mean and unit variance.  Both correspond to the Gaussian white-noise processes $\boldsymbol\xi$ and
$\boldsymbol\eta$ in the Langevin equations [Eq.~\eqref{EqLangevinEquations}].
The three transport coefficients are not independent and we use the relations $D_\text{rot}^0 = 12 D_\perp^0/L^2$ and
$D_\parallel^0 = 2D_\perp^0$ for a slender rod derived within hydrodynamics~\cite{Doi:Oxford:1999}.

\begin{figure*} 
\includegraphics[scale=0.67]{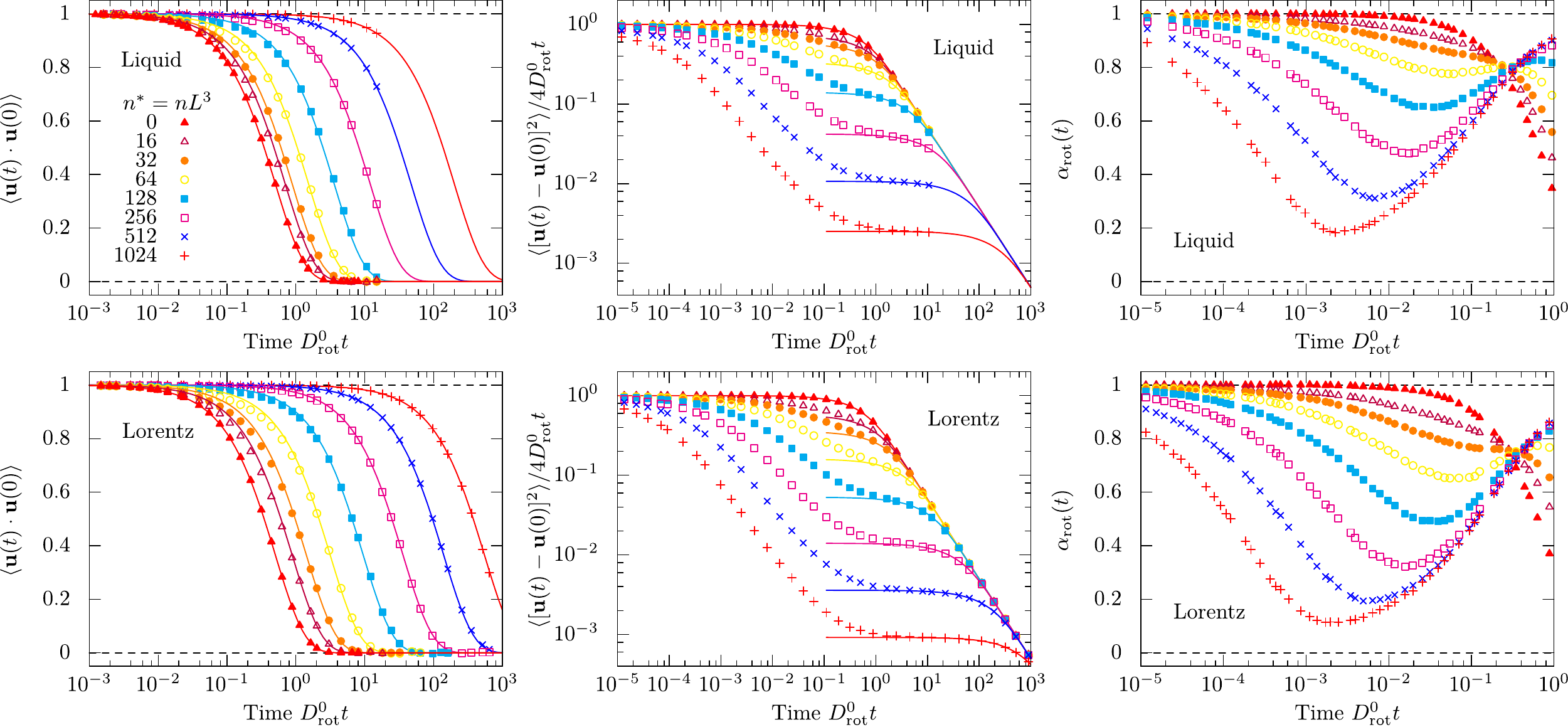}
\caption{\label{FigDACFLinLog}\label{FigDACF}\label{FigLocalExpRot}
Rotational diffusion of the needle characterized by the 
correlation function $\langle \vec{u}(t)\cdot\vec{u}(0)\rangle$ for the orientation $\vec{u}(t)$ (left), 
squared deviation $\langle [\vec{u}(t) - \vec{u}(0)]^2\rangle = 2 - 2\langle\vec{u}(t)\cdot\vec{u}(0)\rangle$ in units
of the short-time rotational motion $4D_\text{rot}^0t$ (middle),
and the local exponent $\alpha_\text{rot}(t)$ (right).
Symbols correspond to simulation results and solid lines represent the phantom needle with long-time rotational diffusion
coefficient $D_\text{rot}$. 
}
\end{figure*}

\subsection{Solution of needles}

In the presence of other needles, the needles can no longer diffuse freely since they are not allowed to cross each
other. To handle these constraints in the simulation, we employ a pseudo-Brownian scheme~\cite{Scala:JCP_126:2007,
Hoefling:JCP_128:2008} to account for the hard-core interaction between the needles.  The detailed steps for the
interaction are outlined in Appendix~\ref{Appendix:HardCoreInteraction}.  Hydrodynamic interactions are ignored, since
they are expected to become unimportant for high aspect ratios~\cite{Dhont:CSA_213:2003}. 

In summary, we subsequently move a single needle and determine possible collisions with other needles during every Brownian
time step of duration $\tau_\text{B}$. Upon collision, we enforce conservation of energy, momentum, and angular
momentum, and the resulting transfer of momentum is directed perpendicular to both orientations of the collision partners (smooth
needles). Special care has been taken that at collisions the flow of energy between the rotational and translational
degrees of freedom of the moving needle vanishes on average, see Appendix~\ref{Appendix:HardCoreInteraction}. 

For needle liquids, we use a Brownian time step of $\tau_\text{B} = 10^{-6}L^2/D_\perp^0$ which is a compromise between
choosing very small times to mimic Brownian motion and choosing larger time steps to reach sufficiently long times. In
the case of the needle Lorentz system, we move a single tracer needle in a quenched array of other needles with the same
hard-core interaction and use $\tau_\text{B} = 10^{-8}L^2/D_\perp^0$. In both cases, we consider the same densities
ranging from infinite dilution to systems deep in the semidilute regime with (reduced) densities over $n^*= nL^3 = 10^3$
where $n$ is the number of needles per volume.

A single configuration for the needle liquid in a simulation box of size $1.5L$ over $10^6$ Brownian time steps takes
around $225$ CPU hours (Intel\textregistered Core\texttrademark i7-4770S @ 3.10GHz) for the highest density and we average
over at least ten realizations. For the needle Lorentz system we use a simulation box of size $25L$ and simulate
over $10^{10}$ Brownian time steps which takes around $35$ CPU hours and we average over $10^3$ trajectories for 
densities $n^* \geq 128$.

\subsection{Phantom needle}

For the needles in solution, the short-time dynamics is described by the short-time diffusion coefficients
$D_\parallel^0$, $D_\perp^0$, and $D_\text{rot}^0$, and differs from the behavior at long times which is characterized
by new transport coefficients for translation, $D_\parallel$ and $D_\perp$, and rotation, $D_\text{rot}$. It is instructive
to compare the dynamics of a needle in solution to that of a single needle with the emerging long-time diffusion
coefficients as input parameters (phantom needle). In computer simulations, this can be easily achieved by evolving the
single needle by the Langevin equations [Eq.~\eqref{EqLangevinEquations}] with the new transport coefficients for
rotation and translation. 

An analytic description is obtained by considering the translationally invariant needle dynamics in space and time in
terms of the propagator $G(\vec{r}, \vec{u}, t|\vec{u}_0)$.  It describes the conditional probability for a displacement
$\vec{r}$ and a change of orientation from $\vec{u}_0$ to $\vec{u}$ of the needle in lag time $t$.  The propagator
$G\equiv G(\vec{r}, \vec{u}, t|\vec{u}_0)$ obeys the initial condition $G(\vec{r}, \vec{u}, t = 0|\vec{u}_0) =
\delta(\vec{r})\delta(\vec{u},\vec{u}_0)$ such that the needle is oriented along $\vec{u}_0$ initially. The
time evolution of the propagator is then determined by the Smoluchowski-Perrin equation~\cite{Doi:Oxford:1999, Berne:Dover:2000}:
\begin{align} \label{EqSmoluchowskiDParaDD} 
\begin{split}
  \partial_t G = D_\text{rot}&\mathcal{R}\cdot(\mathcal{R}G) \\
&+ \partial_\vec{r}\cdot[D_\parallel(\partial_\vec{r} G) - \Delta D (1 - \vec{uu})(\partial_\vec{r} G)]. 
\end{split}
\end{align} The first term on the right hand side contains the rotational operator $\mathcal{R} =
\vec{u}\times\partial_\vec{u}$ and accounts for the change of orientation of the needle in terms of the
rotational diffusion coefficient $D_\text{rot}$.
The second contribution describes the translational diffusion of the needle and couples the diffusional anisotropy
$\Delta D = D_\parallel - D_\perp$ determined by the parallel $D_\parallel$ and the perpendicular diffusion coefficient
$D_\perp$ to the current orientation $\vec{u}(t)$ via the projector $\vec{uu}$. 

Analytic progress is achieved by considering the spatial Fourier transform
$G_\vec{k}(\vec{u}, t | \vec{u}_0) = \int\diff^3r\ e^{-\img \vec{k}\cdot\vec{r}} G(\vec{r}, \vec{u}, t| \vec{u}_0)$,
which fulfills the following equation:
\begin{align}
\begin{split}
  \partial_t G_\vec{k} = D_\text{rot}&\mathcal{R}\cdot(\mathcal{R}G_\vec{k}) \\
&- \{k^2 D_\parallel - \Delta D [k^2 - (\vec{k}\cdot\vec{u})^2]\}G_\vec{k} , 
\end{split}
\end{align}
with wave vector $\vec{k}$ and magnitude $k = |\vec{k}|$.
We choose a representation in spherical coordinates with polar angle $\theta$ and azimuthal angle $\phi$ and fix the
wave vector $\vec{k} = k \vec{e}_z$ along the $z$-direction and we abbreviate $z = \vec{k}\cdot\vec{u}/k = \cos(\theta)$. Then, the
product $\mathcal{R}\cdot\mathcal{R}$ of the rotational operators reduces to the angular momentum operator 
$\mathcal{R}^2_{z,\phi} = \partial_z[(1-z^2)\partial_z] + {(1 - z^2)^{-1}\partial_\phi^2}$ and we obtain
\begin{align} \label{EqSmoluchowskiZPhi}
 \partial_t G_k&= D_\text{rot} \mathcal{R}^2_{z,\phi} G_k - k^2 [D_\parallel - \Delta D (1 - z^2)]G_k.
\end{align}

This partial differential equation is solved by a separation of variables and the full
solution~\cite{Aragon:JCP_82:1985} is obtained as an expansion in terms of spheroidal wave functions $\Ps_n^m$ of degree
$n$ and order $m$:
\begin{align} 
\begin{split} \label{EqPropagatorPhantom}
G_k(z, \phi, t| &z_0, \phi_0) = \sum_{m=-\infty}^\infty \sum_{n = m}^\infty \frac{2n + 1}{4\pi}\frac{(n - m)!}{(n
+ m)!}\times \\ 
&\times\Ps_n^m(z,\gamma^2)\Ps_n^m(z_0,\gamma^2) e^{\img m (\phi -
\phi_0)} e^{-\Gamma_n^m t},
\end{split}
\end{align}
with real parameter $\gamma^2 := k^2\Delta D/D_\text{rot} \geq 0$ and characteristic decay constants $\Gamma_n^m \equiv
\Gamma_n^m(\gamma^2) \geq 0$.  The exact relations for both parameters are determined as solutions of the spheroidal wave
equation~\cite{NIST:DLMF,Olver:2010:NHMF}
\begin{align}
\begin{split}
  \partial_z[(1&-z^2)\partial_z \Ps_n^m] \\ &+ \bigg[\lambda_n^m(\gamma^2) + \gamma^2(1 - z^2) -
\frac{m^2}{1-z^2}\bigg] \Ps_n^m = 0,
\end{split}
\end{align}
with spheroidal eigenvalue $\lambda_n^m \equiv \lambda_n^m(\gamma^2)$. The decay constants are related to the spheroidal
eigenvalue via $\Gamma_n^m = D_\parallel k^2 + D_\text{rot} \lambda_n^m$. 

The propagator $G_k$ [Eq.~\eqref{EqPropagatorPhantom}] contains the full information about the dynamics of the phantom
needle and can be used to obtain explicit expressions for the quantities of interest. 

\section{Transport behavior}

\subsection{Rotational diffusion}

We first consider the effect of the dynamic crowding on the time-dependent orientation $\vec{u}(t)$. A simple quantity
which encodes the topological constraints imposed by the neighboring needles is given by the time-dependent
orientational correlation function $\langle \vec{u}(t)\cdot\vec{u}(0)\rangle$. The angle brackets denote an ensemble average over
all moving needles and configurations and the initial orientation $\vec{u}(0)$ is uniformly distributed over the sphere.

In the absence of other needles, the correlation function decays exponentially, $\langle \vec{u}(t)\cdot\vec{u}(0)\rangle
= \exp(-2 D_\text{rot}^0 t)$, and the time scale for the decay is determined by the rotational diffusion
coefficient $D_\text{rot}^0$~\cite{Doi:Oxford:1999}~[Fig.~\ref{FigDACFLinLog}]. With increasing needle density $n^*$,
the initial orientation $\vec{u}(0)$ persists for longer times since the needle can no longer rotate freely due to the
topological constraints imposed by its neighbors. The time-dependent long-time relaxation of the correlation function is
again captured by an exponential decay $\langle \vec{u}(t)\cdot\vec{u}(0)\rangle = \exp(-2 D_\text{rot} t)$ and the time
scale for the decay encodes the long-time rotational diffusion coefficient $D_\text{rot}$. Deviations from the rotational
motion of the phantom needle are present at intermediate times where the needle explores its close environment and
they become visible for small densities where the concept of a confining tube is not fully applicable.

The time-dependent behavior preceding the exponential decay at long times in the correlation function $\langle
\vec{u}(t)\cdot\vec{u}(0)\rangle$ can be analyzed more closely by the directly related quantity of the squared distance of
the time-dependent orientation $\vec{u}(t)$ and the initial one $\vec{u}(0)$:
$\langle[\vec{u}(t) - \vec{u}(0)]^2\rangle = 2 - 2\langle\vec{u}(t)\cdot\vec{u}(0)\rangle$.
For a freely diffusing needle, the change in orientation is described by ordinary diffusion in two dimensions with diffusion
constant $D_\text{rot}^0$ as long as the needle has not rotated significantly
$(t \ll 1/D_\text{rot}^0)$:
\begin{align} \label{EqShortTimeRotDiff}
\langle[\vec{u}(t) - \vec{u}(0)]^2\rangle = 4 D_\text{rot}^0 t + \mathcal{O}(D_\text{rot}^0 t)^2.
\end{align}

\begin{figure}[t]
\includegraphics[scale=0.85]{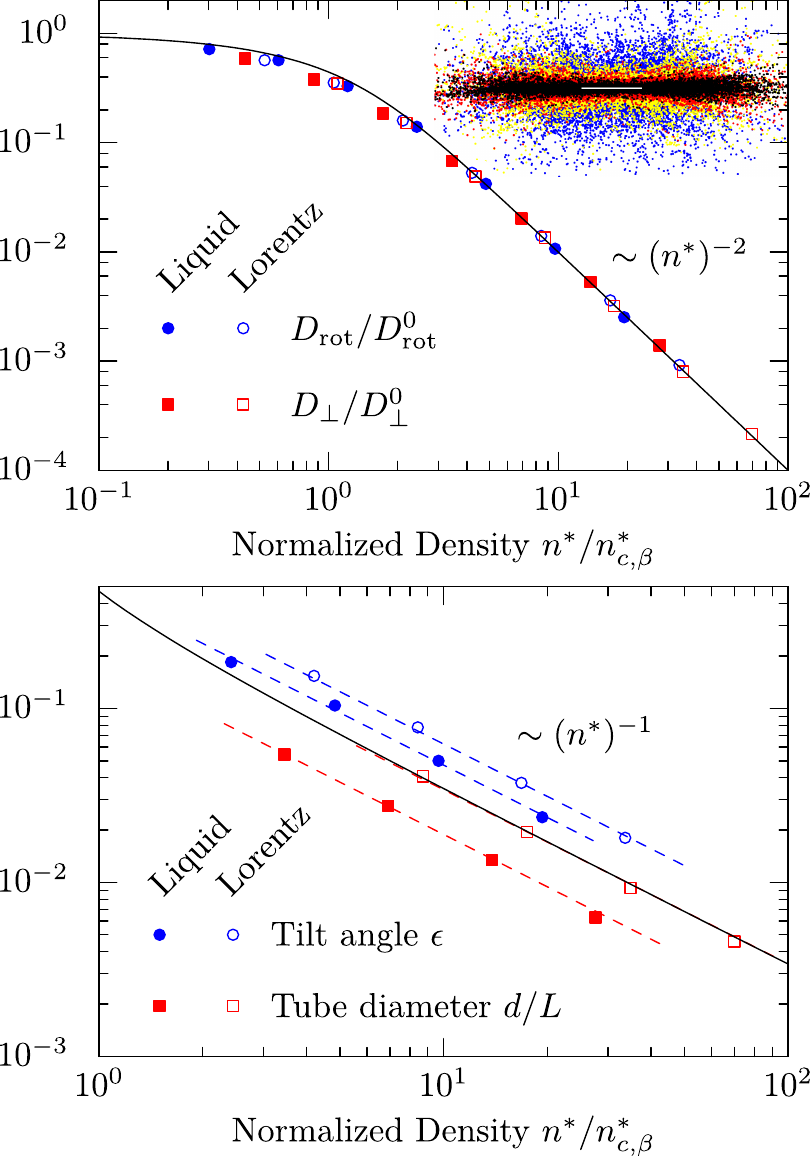}
\caption{\label{FigDiffCoeffTubeScal}
Top: Longtime diffusion coefficients for rotation, $D_\text{rot}$, and perpendicular diffusion $D_\perp$
as a function of the normalized density $n^*/n_{c,\beta}^*$ with respect to the rotational $(\beta = \text{rot})$ and
translational motion $(\beta = \perp)$ for needle liquids as well as needle Lorentz systems. The solid line corresponds
to the theoretical prediction~[Eq.~\eqref{EqSelfConsistentDperp}].  Inset (top): Distribution of the geometric
centers of the needle for densities $n^* = 128$ (blue), $256$ (yellow), $512$ (red), and $1024$ (black) up to times
$L^2/2D_\perp^0$. The initial position and orientation of the needle are indicated by the white needle in the center. For
increasing density, the motion of the needle becomes more and more directed along the initial orientation as anticipated
by the tube theory.  Bottom: Tilt angle $\epsilon$ and tube diameter $d$ with respect to the normalized density.
The solid line represents the theoretical prediction for the localization length~[Eq.~\eqref{EqTubeLocLength}].  Figure
is adjusted from Ref.~\cite{Leitmann:PRL_117:2016}.
}
\end{figure}

\begin{figure*}
\includegraphics[scale=0.85]{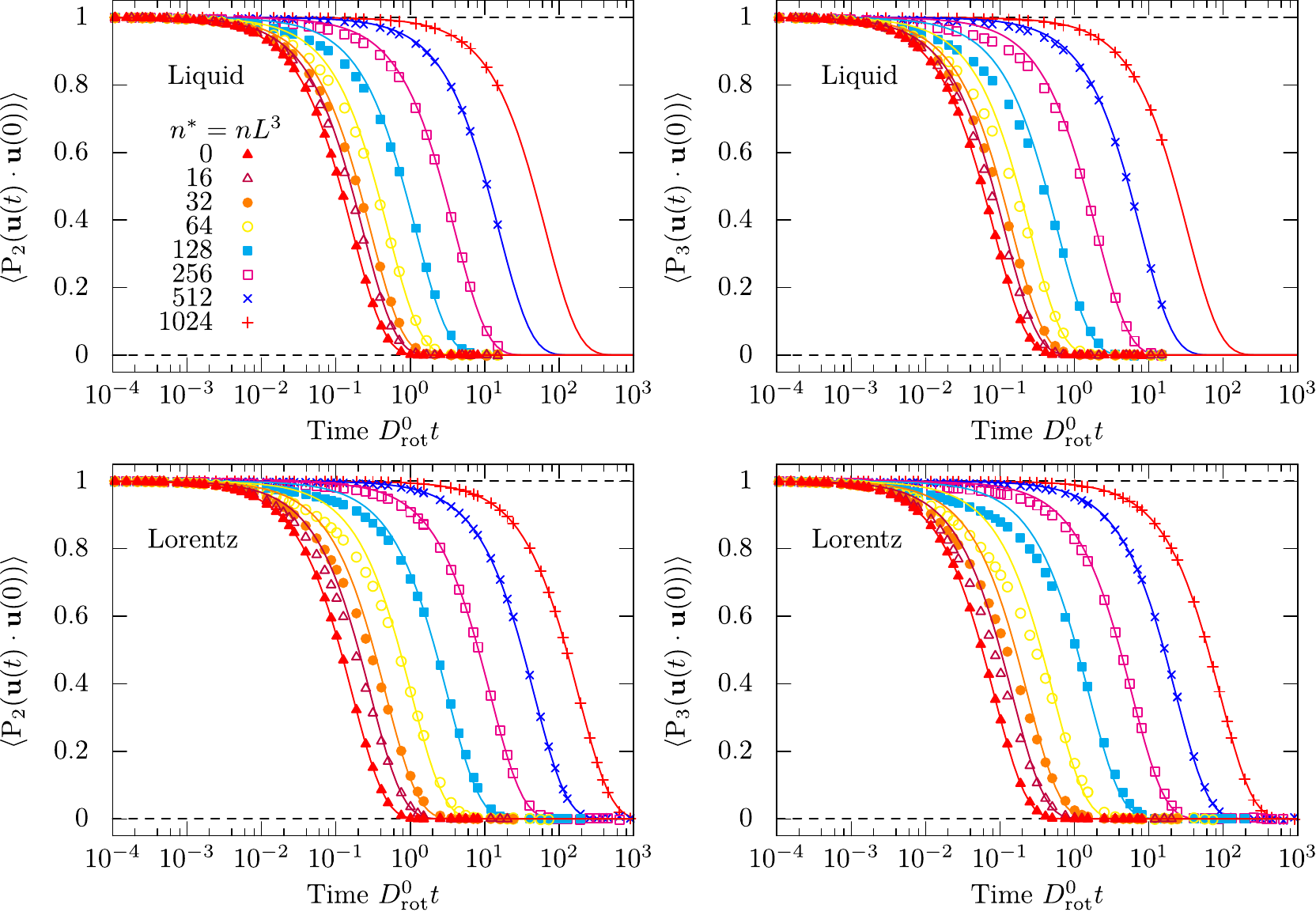}
\caption{\label{FigDACFLinLogP2P3} 
Time-dependent higher-order
orientational correlation functions $\langle \text{P}_\ell\bigl(\vec{u}(t)\cdot\vec{u}(0)\bigr)\rangle$ for $\ell = 2$ (left) 
and $\ell = 3$ (right) at different densities $n^*$ for needle liquids as well as needle Lorentz systems.
Density increases from left to right. Symbols represent results from computer simulations and solid lines represent the
analytic solution $\exp[-\ell(\ell+1)D_\text{rot} t]$ of a freely diffusing phantom needle with long-time rotational diffusion
coefficient $D_\text{rot}$.}
\end{figure*}

\begin{figure*}[t]
\includegraphics[scale=0.66]{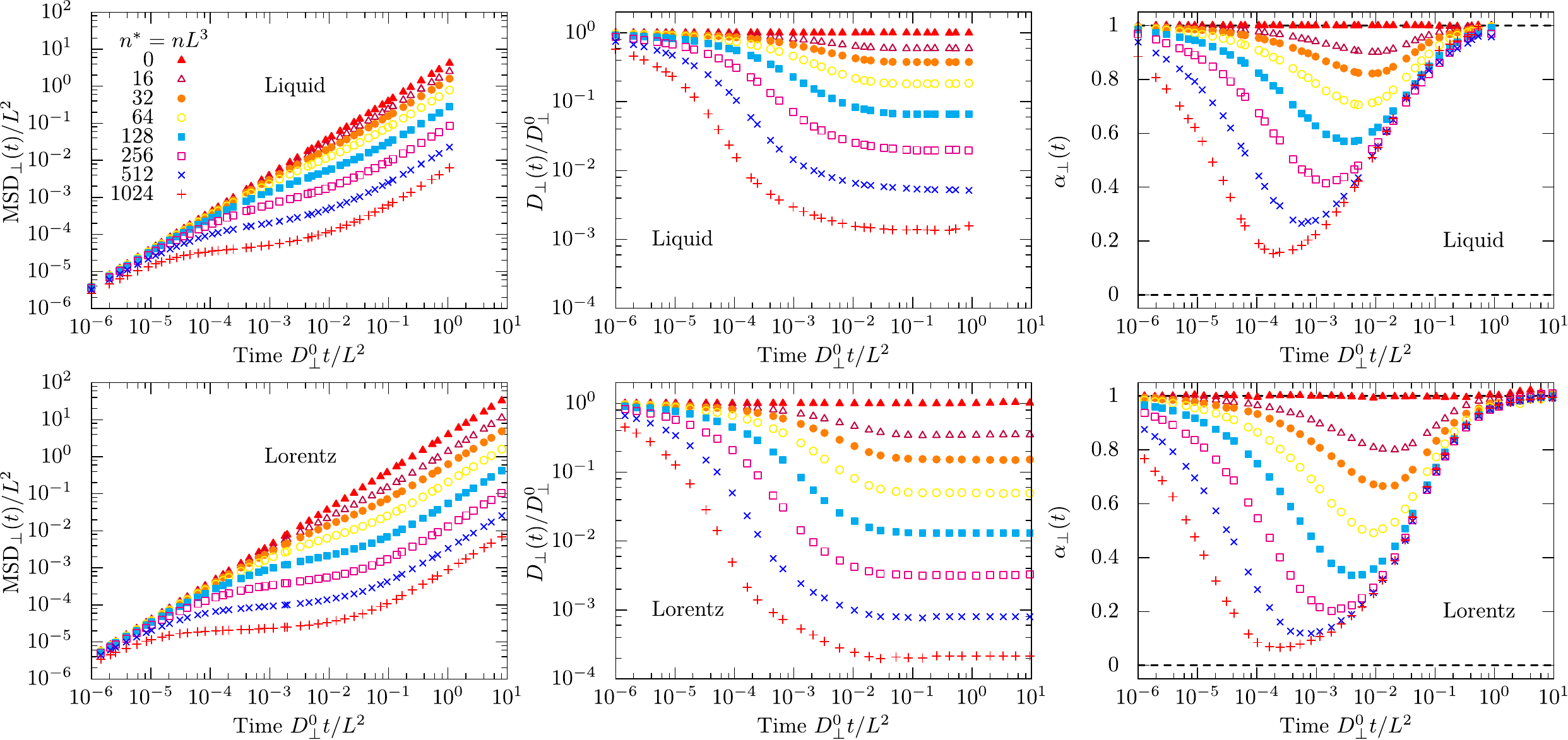}
\caption{\label{FigMSDPerp}\label{FigDiffCoeffPerp}\label{FigLocalExpPerp}
Perpendicular translational diffusion measured in a coordinate frame fixed to the needle characterized by 
the mean-square displacement $\text{MSD}_\perp(t)$ (left), the diffusion coefficient $D_\perp(t)$ (middle), and
the local exponent $\alpha_\perp(t)$ (right). Density increases from top to bottom. Symbols correspond to
simulation results.}
\end{figure*}

The drastic slowing down of rotational diffusion with increasing density of neighboring needles is exemplified by
considering the squared distance of orientations in units of the short-time diffusion of the freely diffusing
needle~[Fig.~\ref{FigDACF}]. For times $t \lesssim 1/D_\text{rot}^0$ the rotational motion
becomes strongly suppressed for high needle densities and approaches
an intermediate plateau at times $t \gtrsim 1/D_\text{rot}^0$ where the squared distance increases linearly with
long-time rotational diffusion coefficient $D_\text{rot}$. This plateau with the following crossover to the exponential
decay of the correlation function offers a direct way to determine $D_\text{rot}$ since it separates the short-time
behavior, where the needle explores its environment, from the emerging stochastic motion of the phantom needle at long
times. The extracted long-time rotational diffusion coefficient follows the scaling law
$D_\text{rot}\sim(n^*)^{-2}$~[Fig.~\ref{FigDiffCoeffTubeScal}] where we normalized the density by the inverse of the
respective prefactor $n_{c,\text{rot}}^* \approx 53$ and $n_{c,\text{rot}}^* \approx 30$ for needle liquids and needle
Lorentz systems, respectively. The scaling behavior has been predicted by theory~\cite{Doi:JP_36:1975,
Teraoka:JCP_91:1989} and also confirmed by earlier computer simulations~\cite{Doi:JPSJ_53:1984, Tao:JCP_124:2006,
Tse:JCP_139:2013}. 

The needle explores its close environment in the time window $t \lesssim 1/D_\text{rot}^0$, which contains information about
the geometry of the confining tube. We measure the suppression of the rotational diffusion by the local exponent 
\begin{align} \label{EqLocalExpRot}
\alpha_\text{rot}(t) = \frac{\diff\ln (\langle[\vec{u}(t) - \vec{u}(0)]^2\rangle)}{\diff\ln(t)} = 
-t \frac{\diff \langle\vec{u}(t)\cdot\vec{u}(0)\rangle/\diff t}{1-\langle \vec{u}(t)\cdot\vec{u}(0)\rangle}.
\end{align}
With increasing density, $\alpha_\text{rot}(t)$ becomes more and more suppressed at intermediate times and we expect a
transient dynamic arrest of the rotational motion by going beyond the densities considered
here~[Fig.~\ref{FigLocalExpRot}].  The time scale for the onset of the relaxation of the arrest is set by the
density-independent time for the needle to diffuse along its long axis $L^2/D_\parallel^0$, leading to a collapse of the
data. For long times, the squared deviation of the orientations on the sphere saturates
${\lim_{t\to\infty}\langle[\vec{u}(t) - \vec{u}(0)]^2\rangle = 2}$ irrespective of the density, resulting in an apparent
common intersection point and a vanishing of the local exponent ${\alpha_\text{rot}(t\to\infty) = 0}$. 

In principle, the tilt angle $\epsilon$ of the needle should become manifest as a plateau of order $\epsilon^2$ in the
deviations of the orientational correlation function from its initial value at intermediate times: $1 - \langle
\vec{u}(t)\cdot\vec{u}(0)\rangle = 1-\cos(\epsilon)$. However, since the plateau is not very well pronounced we use a more
robust determination in terms of the previously defined local exponent for the rotational
motion~[Eq.~\eqref{EqLocalExpRot}]. Here, we define the tilt angle $\epsilon$ via the orientational correlation
function $\langle \vec{u}(\tau_\epsilon)\cdot\vec{u}(0)\rangle = \cos(\epsilon)$ at time $\tau_\epsilon$, at which the
local exponent $\alpha_\perp$ becomes minimal~[Fig.~\ref{FigLocalExpRot}]. In our data such a minimum in the local
exponent $\alpha_\text{rot}$ emerges for times $t < 1/D_\text{rot}^0$ and for densities $n^* \geq 64$ and becomes more
and more pronounced for increasing density. For the tilt angle of the needle in the confining tube, we also recover the
predicted scaling behavior $\epsilon \sim (n^*)^{-1}$~\cite{Doi:Oxford:1999} (Fig.~\ref{FigDiffCoeffTubeScal} reproduced
from Ref.~\cite{Leitmann:PRL_117:2016}).

With the extracted long-time rotational diffusion coefficient $D_\text{rot}$, we can directly compare the dynamics in
needle liquids and needle Lorentz systems to the dynamics of a phantom needle in terms of higher-order orientational
correlation functions
\begin{align} \label{EqHigherDegreeRotCorr}
\bigl\langle\text{P}_\ell\bigl(\vec{u}(t)\cdot\vec{u}(0)\bigr)\bigr\rangle = \exp[-\ell(\ell+1)D_\text{rot}t] .
\end{align}
where $\text{P}_\ell(\cdot)$ denotes the Legendre polynomial of degree $\ell$~\cite{Doi:Oxford:1999}.  
From the full solution of the propagator $G_k$ [Eq.~\eqref{EqPropagatorPhantom}], the preceding relation
[Eq.~\eqref{EqHigherDegreeRotCorr}] can be derived in the following way: For the rotation, we are not
interested in the spatial dynamics and we set $k = 0$ in the expression for the propagator of the phantom needle $G_0
\equiv G_{k=0}$~[Eq.~\eqref{EqPropagatorPhantom}]. Then, the real parameter $\gamma^2 := k^2 \Delta D/D_\text{rot}$
vanishes and the spheroidal wave functions reduce to the associated Legendre polynomials $\Ps_n^m(z,0) =
\text{P}_n^m(z)$ with eigenvalue $\lambda_n^m(0) = n(n+1)$. We express the summands in spherical harmonics
$Y_{nm}(\cdot)$ and obtain with the addition theorem $(2n+1)\text{P}_n(\vec{u}\cdot\vec{u}_0) = 4\pi\sum_{m=-n}^n
Y_{nm}(\vec{u}_0)Y_{nm}^*(\vec{u})$ the propagator for the rotational dynamics~\cite{Teraoka:JCP_91:1989}:
\begin{align}
G_0(\vec{u}, t|\vec{u}_0) = \sum_{n = 0}^\infty \frac{2n+1}{4\pi}e^{-n(n+1)D_\text{rot} t} \text{P}_n(\vec{u}\cdot\vec{u}_0) .
\end{align}
The correlation functions~[Eq.~\eqref{EqHigherDegreeRotCorr}] then follow by averaging over all initial
orientations and integrating over all final orientations: 
\begin{align}
\langle\text{P}_\ell\bigl(\vec{u}(t)\cdot\vec{u}(0)\bigr)\rangle
&= \int\diff^2 u\int\frac{\diff^2 u_0}{4\pi}
\text{P}_\ell(\vec{u}\cdot\vec{u}_0)G_0(\vec{u},t|\vec{u}_0).
\end{align}

The measured correlation functions of orders two and three show stronger deviations at smaller densities than the
first-order orientational correlation function~[Fig.~\ref{FigDACFLinLogP2P3}]. However, for high entanglement, the
dynamics of the phantom needle and the needle liquid and needle Lorentz systems are indistinguishable. Thus, on
coarse-grained time scales and for the strong entanglement the pure orientational motion is described by simple
diffusion on a sphere.

\subsection{Translational diffusion}

We first discuss the translational dynamics in a coordinate frame comoving with the needle. The coordinate frame is
defined by three orthonormal basis vectors: one aligned parallel to the long axis, $\vec{u}(t)$, and two perpendicular directions denoted by
$\vec{u}_1(t)$ and $\vec{u}_2(t)$. 
The time evolution of $\vec{u}(t)$ is described in Eq.~\eqref{EqPropagationRules}, whereas the perpendicular directions follow by rotating 
\begin{align}
\begin{split}
\vec{u}_i(t + \Delta t) = &\vec{u}_i(t)\cos(|\boldsymbol\omega|\Delta t) +
\biggl(\frac{\boldsymbol\omega}{|\boldsymbol\omega|}\times \vec{u}_i(t)\biggr)\sin(|\boldsymbol\omega|\Delta t) \\
&+ \frac{\boldsymbol\omega}{|\boldsymbol\omega|}\biggl(\frac{\boldsymbol\omega}{|\boldsymbol\omega|}\cdot
\vec{u}_i(t)\biggr)[1-\cos(|\boldsymbol\omega|\Delta t)].
\end{split}
\end{align}
Then, the parallel and perpendicular
displacements in the comoving frame with respect to the displacement of the geometric center
are obtained by $\Delta r_\parallel(t) = \int_0^t\diff t'\ \vec{v}(t')\cdot\vec{u}(t')$ and $
\Delta r_i(t) = \int_0^t\diff t'\ \vec{v}(t')\cdot\vec{u}_i(t')$, respectively.

\begin{figure*}[t]
\includegraphics[scale=0.67]{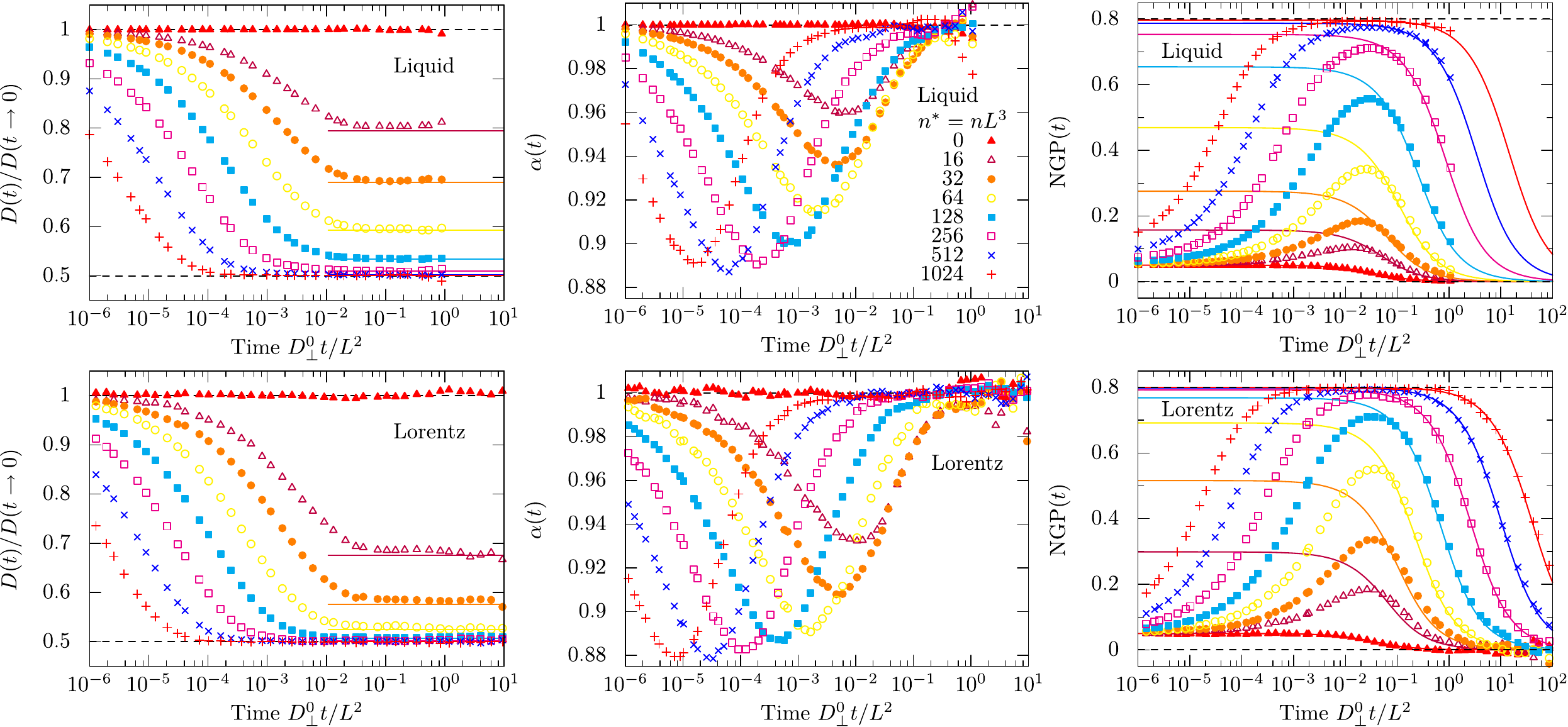}
\caption{\label{FigMSD}\label{FigNonGaussian}
Time-dependent diffusion coefficient of the geometric center $D(t)$ (left), the local exponent $\alpha(t)$ (middle), 
and non-Gaussian parameter $\text{NGP}(t)$ (right) at different densities for needle liquids as well as
needle Lorentz systems. Symbols correspond to simulation results and lines represent the behavior of the phantom needle.
}
\end{figure*}

Since we consider infinitely thin smooth needles, the collisions do not influence the parallel motion and the
mean-square displacement parallel to the long axis, $\text{MSD}_\parallel(t) = \langle \Delta r_\parallel^2(t)\rangle$,
is unaffected by the dynamic crowding and it is diffusive at all times with long-time diffusion coefficient
$D_\parallel=D_\parallel^0$:
\begin{align}
\text{MSD}_\parallel(t) =  2 D_\parallel^0 t.
\end{align}
The mean-square displacement perpendicular to the long axis, $\text{MSD}_\perp(t) =\langle \Delta r^2_1(t)\rangle +
\langle\Delta r^2_2(t)\rangle$, shows a strong suppression with increasing density
at intermediate times due to the caging by neighboring needles with a crossover to ordinary diffusion for long times~[Fig.~\ref{FigMSDPerp}]. 
We define the time-dependent diffusion coefficient perpendicular to the needle by the derivative
\begin{align}
D_\perp(t) = \frac{1}{4}\frac{\diff}{\diff t} \text{MSD}_\perp(t) . 
\end{align}
Then, the regime of ordinary diffusion at long times is reflected by a plateau, which encodes the perpendicular long-time
diffusion coefficient $D_\perp := D_\perp(t\to \infty)$~[Fig.~\ref{FigDiffCoeffPerp}]. The extracted diffusion coefficient $D_\perp$ exhibits
the same scaling behavior~[Fig.~\ref{FigDiffCoeffTubeScal}] as the rotational long-time diffusion coefficient,
$D_\perp\sim (n^*)^{-2}$, and the scaling behavior also confirms the theoretical
predictions~\cite{Teraoka:JCP_89:1988,Szamel:PRL_70:1993,Szamel:JCP_100:1994}.  For the normalization of the density we
used the square root of the prefactor of the scaling behavior, which reads $n_{\text{c},\perp}^* \approx 37$ and
$n_{\text{c},\perp}^* \approx 15$ for needle liquids and needle Lorentz systems, respectively.  The
density-dependent behavior of the diffusion coefficient $D_\perp$ is captured by a theoretical prediction for solutions
of nonrotating infinitely thin needles~\cite{Szamel:PRL_70:1993,Szamel:JCP_100:1994} where the suppression of the
diffusion coefficient can be expressed in terms a self-consistent equation of the form 
\begin{align} \label{EqSelfConsistentDperp}
D_\perp/D_\perp^0 = \frac{1}{1+(D_\perp^0/2D_\perp)^{1/2}n^*\lambda_\perp(D_\perp/2D_\perp^0)}, 
\end{align}
where $\lambda_\perp(\cdot)$ denotes a complicated monotonic function and we used $D_\parallel^0 = 2 D_\perp^0$. An
explicit expression of $\lambda_\perp(\cdot)$ is given in Eq. (A$16$) of Ref.~\cite{Szamel:JCP_100:1994}. For strong
suppression, the asymptotic behavior of the self-consistent equation [Eq.~\eqref{EqSelfConsistentDperp}] is obtained as
$D_\perp/D_\perp^0 \sim 36\pi (n^*)^{-2}$, which defines the normalization $n^*_{c,\perp} = 6\sqrt{\pi}$ for comparison
with our simulation results.

Information about the geometry of the confining tube is contained in the time-dependent suppression of the perpendicular
mean-square displacement $\text{MSD}_\perp(t)$ and can be extracted by considering the local exponent 
\begin{align} \label{EqLocalExpPerp}
\alpha_\perp(t) = \frac{\diff \ln(\text{MSD}_\perp(t))}{\diff \ln(t)} = \frac{2D_\perp(t) t}{\text{MSD}_\perp(t)},
\end{align}
similar to the tilt angle for the rotational motion~[Eq.~\eqref{EqLocalExpRot}]. With increasing density, the local exponent becomes 
more and more suppressed at intermediate times and nearly vanishes for the highest density considered in the Lorentz
system. Similar to the local exponent for the rotation $\alpha_\text{rot}(t)$~[Eq.~\eqref{EqLocalExpRot}], we expect a transient
dynamic arrest at intermediate times by further increasing the density. The time scale for the increase of the local
exponent is again set by the time scale for the parallel diffusion leading to a collapse of the data for high
entanglement. We use the time $\tau_\perp$ corresponding to the minimum in $\alpha_\perp(t)$ to define the tube diameter
$d$ via $\text{MSD}_\perp(\tau_\perp) = d^2$, which exhibits the predicted scaling,
$d\sim(n^*)^{-1}$~\cite{Szamel:PRL_70:1993,Sussman:PRE_83:2011,Sussman:PRL_107:2011}~[Fig.~\ref{FigDiffCoeffTubeScal}]. 
We also compare the tube diameter $d$ obtained from our simulations to a theoretical prediction of the tube localization
length for nonrotating infinitely thin needles. Just recently, the density-dependent behavior of the tube-localization
length $r_\text{loc}$ has been extended to all densities~\cite{Sussman:PRE_83:2011,Sussman:PRL_107:2011} and is given by
the self-consistent equation
\begin{align} \label{EqTubeLocLength}
\frac{L^2}{r_\text{loc}^2} = \frac{\pi n^*}{4\sqrt{2}}\frac{L^2}{r_\text{loc}^2}\lambda_\text{loc}(L/r_\text{loc}),
\end{align}
for nonrotating needles performing only perpendicular diffusion. The function $\lambda_\text{loc}(x) = [x - I_1(2x) +
L_1(2x)]/2x^2$ is defined in terms of the first modified Bessel function $I_1(\cdot)$ and the first modified Struve function
$L_1(\cdot)$.  For very small localization lengths $r_\text{loc}/L \ll 1$, the self-consistent equation reduces to the
previously known scaling behavior $r_\text{loc}/L \sim 8\sqrt{2}/\pi n^*$~\cite{Szamel:PRL_70:1993,Szamel:JCP_100:1994}. 
The predicted tube-localization length with normalization
$n_{c,\perp}^* = 6\sqrt{\pi}$ nicely captures the range of tube diameters $d$ for needle liquids and needle
Lorentz systems~[Fig.~\ref{FigDiffCoeffTubeScal}]. Interestingly, the prediction coincides with our result of the tube
diameter for the needle Lorentz system. 

\begin{figure*}
\includegraphics[width=0.85\linewidth]{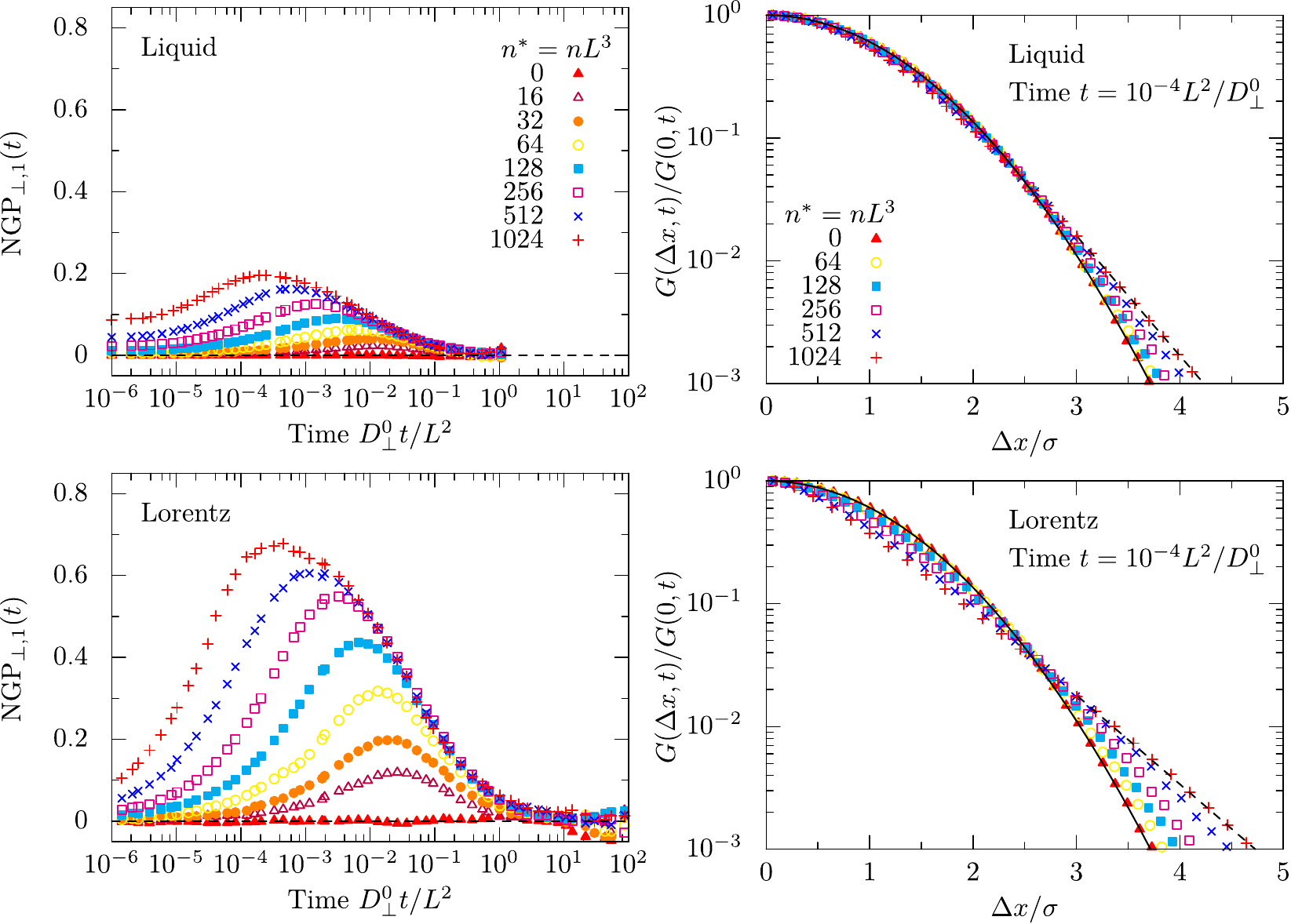}
\caption{\label{FigVanHove}
Left: Non-Gaussian parameter in the comoving frame along a direction perpendicular to the needle axis for needle liquids
as well as needle Lorentz systems. Right: Probability distribution $G(\Delta x, t)$ of the displacements $\Delta x$ at
time $t = 10^{-4} L^2/D_\perp^0$ along one direction perpendicular to the needle axis measured in the comoving frame.
Symbols correspond to computer simulations. The displacement $\Delta x$ is normalized by the standard deviation $\sigma$
of the respective distribution $G(\Delta x, t)$. The solid line represents the Gaussian behavior, and the dashed lined
indicates an exponential decay for large displacements.
}
\end{figure*}

The next interesting quantity for the translational diffusion of the needle is given by the mean-square displacement
of the geometric center in the laboratory fixed frame, $\text{MSD}(t) = \langle |\Delta\vec{r}(t)|^2\rangle$. 
We discuss the time dependence of the mean-square displacement via the time-dependent diffusion coefficient
[Fig.~\ref{FigMSD}]
\begin{align}
D(t) = \frac{1}{6}\frac{\diff}{\diff t}\text{MSD}(t).
\end{align}

For short times, the diffusion coefficient is determined by the short-time diffusion coefficients for translation:
$D(t\to 0) = (D_\parallel^0 + 2 D_\perp^0)/3$. In the presence of other needles, the diffusion coefficient decreases
over time due to the suppression of the perpendicular needle motion~[Fig.~\ref{FigMSD}].  For strong entanglement, the
contributions from translational diffusion perpendicular to the needle vanish in comparison to the parallel one, and the
long-time diffusion coefficient $D := D(t\to \infty)$ approaches its limiting value $D_\parallel/3$.  On coarse-grained time scales, the phantom
needle again captures the dynamics.

The translational dynamics of the geometric center can also be discussed in terms of the local exponent
\begin{align}
\alpha(t) = \frac{\diff \ln(\text{MSD}(t))}{\diff \ln(t)} = \frac{2 D(t) t}{\text{MSD}(t)}.
\end{align}
For increasing density, the crossover regime shifts to earlier times~[Fig.~\ref{FigNonGaussian}] as anticipated
from the diffusion coefficient $D(t)$. The local exponent exhibits a lower bound for the considered densities and by
scaling the time with the squared density, the data collapse for densities $n^* \gtrsim 256$.  Since the parallel
diffusion of the needle is unaffected by neighboring needles, the time-dependent behavior of the local exponent
$\alpha(t)$ is quite different from that found in models of porous
media~\cite{Skinner:PRL_111:2013,Schnyder:SM_11:2015,Spanner:PRL_116:2016} and glass- forming
systems~\cite{Voigtmann:PRL_103:2009, Horbach:EPJST_141:2010, Goetze:Oxford:2012}, which exhibit anomalous diffusion and a
localization transition.

Deviations from ordinary diffusion are quantified by the non-Gaussian parameter defined
via~\cite{Rahman:PR_136:1964,Hoefling:RPP_76:2013}
\begin{align}
\text{NGP}(t) = \frac{3}{5}\frac{\text{MQD}(t)}{\text{MSD}(t)^2}  - 1, 
\end{align}
with mean-quartic displacement $\text{MQD}(t) = \langle |\Delta \vec{r}(t)|^4\rangle$ of the geometric needle center.
For the phantom needle it follows via a time-dependent perturbation theory with
Eq.~\eqref{EqSmoluchowskiZPhi}~\cite{Dibak:Bachelor:2013} (for a derivation in terms of spheroidal wave functions
see Ref.~\cite{Kurzthaler:SR_6:2016} and Appendix~\ref{Appendix:Moments}):
\begin{align}
\text{MQD}(t) = 60 D^2 t^2 + \frac{8(\Delta D)^2}{27 D_\text{rot}^2}[6 D_\text{rot} t - 1 + e^{-6 D_\text{rot} t}].
\end{align}
Thus, for the non-Gaussian parameter of the phantom needle, it follows that
\begin{align} \label{EqNonGaussianParameter}
\text{NGP}(t) = \frac{8}{45}\biggl(\frac{\Delta D/D}{6 D_\text{rot} t}\biggr)^2[6 D_\text{rot} t - 1 + e^{-6 D_\text{rot} t}].
\end{align}
In particular, $\text{NGP}(t)$ is nonvanishing for anisotropic diffusion and it approaches zero only algebraically
$\text{NGP}(t) = \mathcal{O}(t^{-1})$ as anticipated by the central limit theorem. 

In the presence of other needles the short-time non-Gaussian parameter is determined by the short-time diffusion
coefficients and is given by $\text{NGP}(t\to 0) = 1/20$ for a slender rod. With increasing density of needles,
the rotational and the translational diffusion coefficient perpendicular to the needle axis become more
and more suppressed. In particular, the perpendicular diffusion coefficient becomes negligible in comparison to the
parallel one which is not affected by the disorder. Thus, for the phantom needle, the diffusional anisotropy $\Delta D$ and
diffusion coefficient $D$ are purely determined by the parallel diffusion
coefficient with $\Delta D = D_\parallel$ and $D = D_\parallel/3$. In this case
and for times where the rotational motion is small, the non-Gaussian parameter
[Eq.~\eqref{EqNonGaussianParameter}] of the phantom needle reduces to 
\begin{align}
\text{NGP}(t) = 
\frac{4}{5}[1 - 2 D_\text{rot} t + \mathcal{O}(D_\text{rot} t)^2], 
\end{align}
and the purely translational motion parallel to the long axis of the phantom needle emerges as a plateau $4/5$ as long as
the phantom needle has not rotated significantly. This plateau is observed in our data for the highest densities considered 
for the needle in solution~[Fig.~\ref{FigNonGaussian}]. There, the needle probes its confining tube of diameter $d$ at time scale
$d^2/D_\perp^0\sim 1/(n^*)^2$, whereas the long-time rotational relaxation becomes relevant only for times
$1/D_\text{rot} \sim (n^*)^2$.  
Again, we find that the phantom needle captures the time-dependent behavior at high densities and on coarse-grained time
scales, once the needle has explored its initial tube and rotates with long-time diffusion coefficient $D_\text{rot}$.

The contributions to the $\text{NGP}(t)$ discussed above arise essentially from the anisotropy of the diffusion, masking
actual non-Gaussian behavior in the individual Cartesian components. Since the needle can freely diffuse along its long
axis irrespective of the density of neighboring needles, the non-Gaussian parameter of $\Delta r_\parallel(t)$, measured
in the comoving frame along the needle axis vanishes. This is in striking contrast to the motion perpendicular to the
needle axis which exhibits strong deviations from ordinary diffusion as indicated by the non-Gaussian parameter
$\text{NGP}_{\perp,1}(t)$ of the displacement $\Delta x\equiv \Delta r_1$ of the needle perpendicular to its long axis
along one direction in the comoving frame~[Fig.~\ref{FigVanHove} (left)]. Here, we further investigate this
scenario in terms of the probability distribution $G(\Delta x, t)$ of the displacements $\Delta x\equiv \Delta r_1$,
which we evaluate at time $t = 10^{-4} L^2/D_\perp^0$ to minimize the influence of the parallel motion of the needle as
indicated by the local exponent $\alpha_\perp(t)$~[Fig.~\ref{FigLocalExpPerp} (right)]. Hence, the needle only
probes the confining tube due to the neighboring needles. 

The probability distribution $G(\Delta x,t)$ develops significant deviations from a Gaussian with increasing density of
the needles~[Fig.~\ref{FigVanHove} (right)].  In
particular for the needle Lorentz system and for large displacements, the distribution $G(\Delta x, t)$ approaches an
exponential, which may be interpreted as a constant effective restoring force $f = - k_\text{B} T \diff \ln[G(\Delta x, t)]/\diff
(\Delta x)$ on the needle due to the confining tube. This scenario has been anticipated theoretically just recently for
solutions of infinitely thin needles where the full tube confinement potential has been constructed for the first
time~\cite{Sussman:PRL_107:2011}. The tube confinement potential exhibits a strongly anharmonic character and thus a
distribution of tube diameters or transverse localization lengths on intermediate scales. Experimentally, this scenario
has also been observed for entangled solutions of semiflexible polymers where the tube width distribution is not
harmonic due to the existence of stretched tails and a displacement-independent effective restoring
force~\cite{Wang:PRL_104:2010,Glaser:PRL_105:2010}.  While these anharmonic displacements are also observed for needle
liquids, they are much less pronounced in comparison to needle Lorentz systems due to the existence of constraint
release processes. 

Such exponential distributions of the displacement are also found in glassy materials~\cite{Chaudhuri:PRL_99:2007}.
However, we have to stress that both situations are different since the needle can always diffuse along its long axis
irrespective of the density and is only confined perpendicular to its long axis.

\subsection{Intermediate scattering function of the needle}

Information about the spatiotemporal dynamics of the needle center is encoded in the intermediate scattering function 
\begin{align} \label{EqIntScatNeedleCenter}
F(k, t) = \langle e^{-\img\vec{k}\cdot\Delta\vec{r}(t)}\rangle
=\int\!\diff^2 u \int\!\frac{\diff^2 u_0}{4\pi} G_k(\vec{u},t|\vec{u}_0),
\end{align}
with wave vector $\vec{k}$ and wave number $k = |\vec{k}|$. Both integrals extend over all possible initial and final
orientations. We have discussed the intermediate scattering function $F(k,t)$ recently
for needle liquids as well as needle Lorentz systems and rationalized the results in terms of the phantom needle~\cite{Leitmann:PRL_117:2016}.

\begin{figure*}[t]
\includegraphics[scale=0.85]{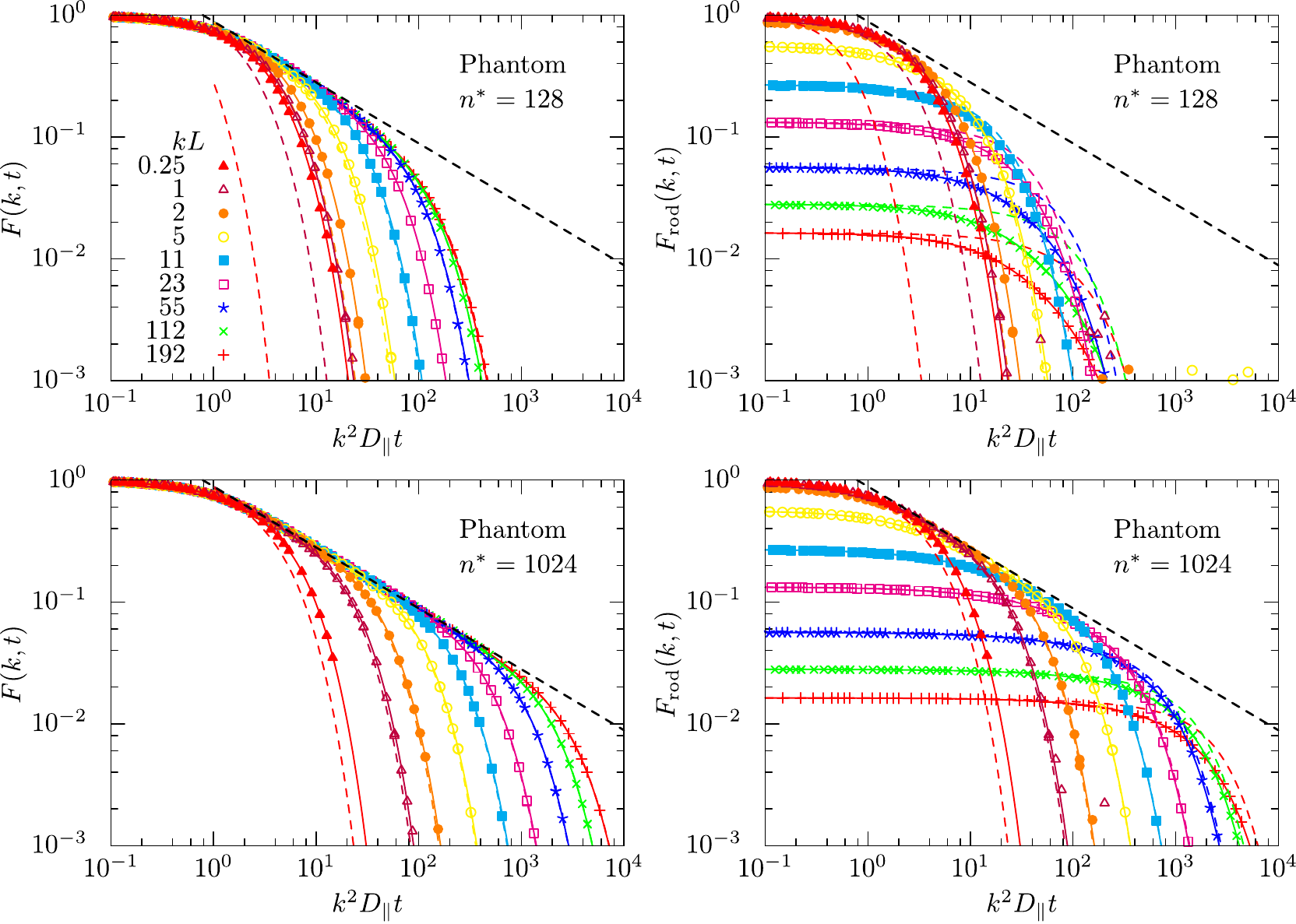}
\caption{\label{FigCompIntScatClosedExp}
Comparison of the intermediate scattering function of the center of mass, $F(k,t)$ (left), and of the whole needle,
$F_\text{rod}(k,t)$ (right), of the phantom needle with transport coefficients from the needle Lorentz systems.
Solid lines represent to the analytic solution and symbols correspond to simulation results of the phantom needle.  The
colored dashed lines represent the approximative solutions
[Eqs.~\eqref{EqIntScatHermiteApprox}~and~\eqref{EqIntScatRodApprox}] and the black dashed lines corresponds to the algebraic
decay $\sim t^{-1/2}$ emerging from the sliding motion in the confining tube.  The left panels are reproduced from Ref.~\cite{Leitmann:PRL_117:2016}.
}
\end{figure*}

\begin{figure*}
\includegraphics[scale=0.85]{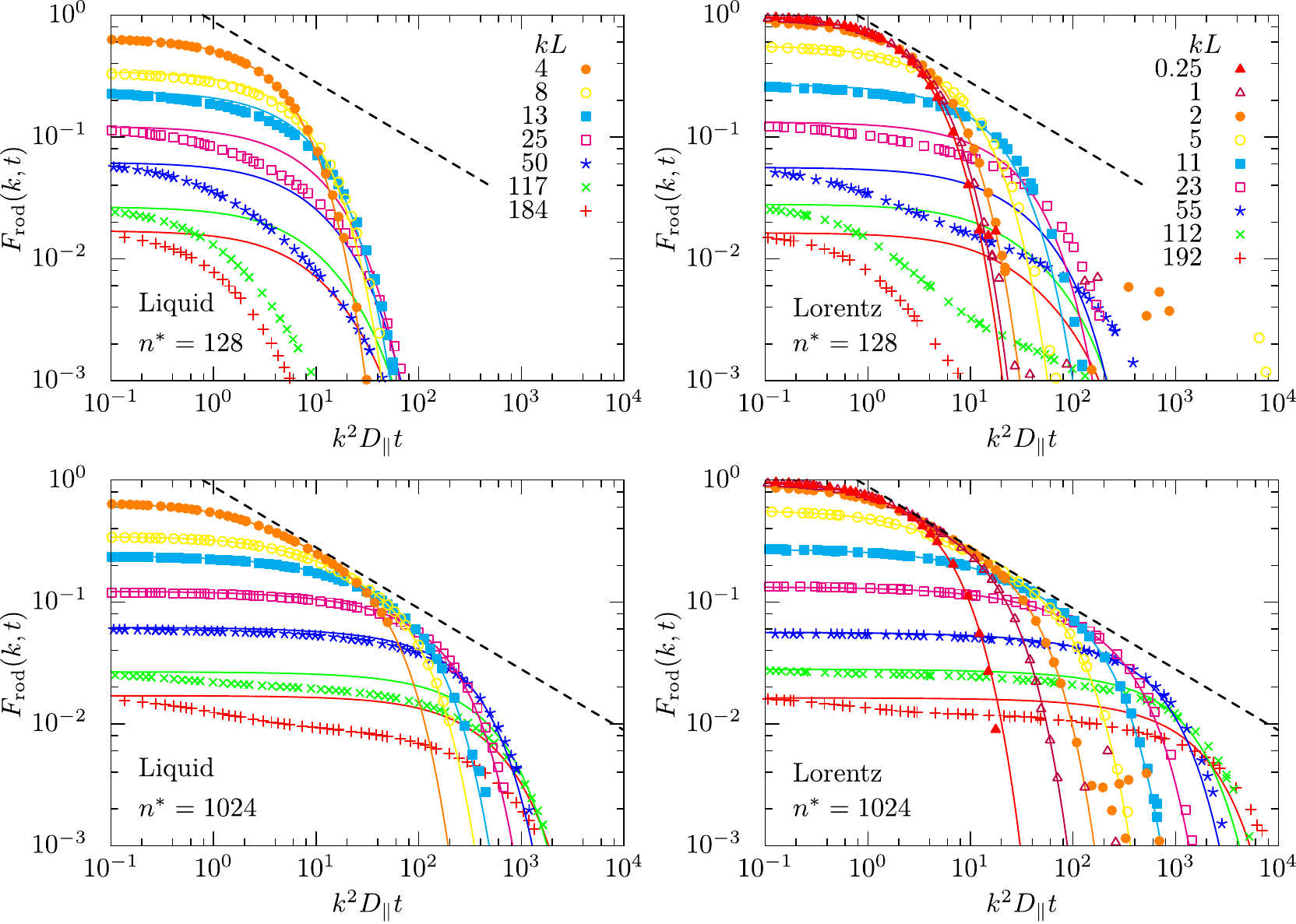}
\caption{\label{FigIntScatRodLiqLor}
Intermediate scattering function of the whole rod $F_\text{rod}(k,t)$ at different densities $n^*$ for needle liquids as well as
needle Lorentz systems. Symbols correspond to simulation results and solid lines represent the analytic solution
[Eq.~\eqref{EqIntScatRodFullSol}] for the phantom needle. The black dashed lines corresponds to the algebraic decay $\sim
t^{-1/2}$ emerging from the sliding motion in the confining tube. Wave number $k$ increases from top to bottom.
}
\end{figure*}

Here, we consider the intermediate scattering function in the case that all segments of the needle contribute to the
scattering. For the two-dimensional analog where the needle moves in a planar array of point obstacles, the intermediate
scattering function for the geometric center and the entire needle has been evaluated earlier and also compared to the
phantom needle~\cite{Munk:Phd:2008,Munk:EPL_85:2009}.

We introduce the fluctuating density of the needle 
\begin{align}
\rho_\text{rod}(\vec{k},t) = \frac{1}{L}\int_{-L/2}^{L/2}\diff s\ \exp(\img\vec{k}\cdot[\vec{r}(t)+s \vec{u}(t)]) ,
\end{align}
where the integral extends over all segments $s$ of the needle. The corresponding intermediate scattering function is obtained via
\begin{align}
F_\text{rod}(k,t) = \langle \rho_\text{rod}(\vec{k},t)^* \rho_\text{rod}(\vec{k},0)\rangle , 
\end{align}
where the angle brackets denote the same average as in the case of the intermediate scattering function of the needle
center~[Eq.~\eqref{EqIntScatNeedleCenter}] and $^*$ is the complex conjugate. 
Thus, the quantity of interest is
\begin{align}
\begin{split}
\label{EqIntScatRodSegments}
F_\text{rod}(k,t) = \frac{1}{L^2}\int\!\diff s \int\!\diff s_0\ \langle e^{-\img\vec{k}\cdot\Delta\vec{r}(t)}
e^{-\img\vec{k}\cdot [s \vec{u}(t) - s_0\vec{u}(0)]}\rangle \\
= \frac{1}{L^2}\int\!\diff s\int\!\diff s_0 \int\!\diff^2 u \int\!\frac{\diff^2 u_0}{4\pi}
e^{-\img\vec{k}\cdot [s \vec{u} - s_0\vec{u}_0]} G_k(\vec{u},t|\vec{u}_0).
\end{split}
\end{align}
We follow the solution strategy in Ref.~\cite{Aragon:JCP_82:1985} and express the exponentials containing the segment of
the needle in terms of the Rayleigh expansion,
\begin{align}
e^{\img k s z} = \sum_{l=0}^\infty (2l+1) \img^l j_l(k s)\text{P}_l(z),
\end{align}
with spherical Bessel function $j_l(\cdot)$ and Legendre polynomial $\text{P}_l(\cdot)$. 
Then, one can perform the average over all initial orientations
$\vec{u}_0$ and integrate over all final orientations $\vec{u}$ with respect to the propagator
$G_k$~[Eq.~\eqref{EqPropagatorPhantom}]. Since the intermediate scattering function $F_\text{rod}(k,t)$ is isotropic,
only spheroidal wave functions of order $m=0$ and even degree $n$ contribute and we expand the remaining spheroidal
wave functions in Legendre polynomials via 
\begin{align} \label{EqExpansionSpheroidal}
\Ps_{n}^0(z, \gamma^2) &= \sum_{k=-\lfloor n/2\rfloor}^\infty (-1)^k a_{n,k}^0(\gamma^2)\text{P}_{n+2k}(z).
\end{align}
Then, we obtain an expression suitable for numerical evaluation:
\begin{align} \label{EqIntScatRodFullSol}
F_\text{rod}(k,t) = \sum_{\substack{n=0\\ n\text{ even}}}^\infty (2n + 1) A_n^2 e^{-\Gamma_n^0 t} .
\end{align}
The coefficients 
\begin{align}
A_n = \sum_{\substack{l=0\\l \text{ even}}}^\infty \bigl(a_{n,-(n-l)/2}^0\bigr) \frac{1}{L} \int_{-L/2}^{L/2} \diff s\ j_l(ks) 
\end{align}
contain higher-order expansion coefficients $a_{n,-(n-l)/2}^0$ of the spheroidal wave
functions~[Eq.~\eqref{EqExpansionSpheroidal}] which can be evaluated
numerically~\cite{Hodge:JMP_11:1970,Falloon:JPA:2003}. Moreover, only even orders $l$ have to be considered due to the
symmetry property $j_l(-ks)=(-1)^l j_l(ks)$ of the spherical Bessel functions.

In computer simulations, we evaluate the intermediate scattering function via the formula
\begin{align}
F_\text{rod}(k,t) 
= \bigl\langle e^{-\img\vec{k}\cdot\Delta\vec{r}(t)}f_\vec{k}\bigl(\vec{u}(t)\bigr)f_\vec{k}\bigl(\vec{u}(0)\bigr)\bigr\rangle ,
\end{align}
with the form factor $f_\vec{k}(\vec{u}) = \sin(\vec{k}\cdot\vec{u}L/2)/(\vec{k}\cdot\vec{u}L/2)$. The preceding
equation is obtained after performing the integration over all segments in Eq.~\eqref{EqIntScatRodSegments}. The
presence of the form factor in $F_\text{rod}(k,t)$ suggests that the scattering signal decorrelates faster than in
$F(k,t)$ by the rotation of the needle.

First, we discuss both intermediate scattering functions $F(k,t)$ and $F_\text{rod}(k,t)$ for the phantom needle with
transport coefficients obtained from the needle Lorentz system at densities $n^*=128$ and
$n^*=1024$~[Fig.~\ref{FigCompIntScatClosedExp}].  The full analytic solution for $F_\text{rod}(k,t)$ is corroborated by the
simulation results for the phantom needle for all wave numbers and all times considered.

For small wave number $kL \lesssim 1$, both $F(k,t)$ and $F_\text{rod}(k,t)$ show similar time-dependent
behavior since the dynamics is dominated by the diffusion of the geometric center of the needle~[Fig.~\ref{FigCompIntScatClosedExp}].
For increasing wave number, differences become apparent already at small times, since the intermediate scattering
function of the geometric center is normalized, $F(k,0) = 1$, while the scattering function of the entire needle
exhibits a static structure, i.e., a wave-number dependent initial value,
$F_\text{rod}(k,0) = \int_{-1}^1 \diff z\ j_0(kLz/2)^2/2 = 1 - (kL)^2/36 + \mathcal{O}(kL)^4$.

At intermediate times, both scattering functions display a characteristic algebraic decay of the form $\sim t^{-1/2}$,
which is a fingerprint of the sliding motion of the needle~\cite{Doi:JCSFT_74:1978}. In particular, the intermediate
scattering function for the geometric center, $F(k,t)$, can be evaluated in closed form for times $D_\text{rot}\gamma^2 t
\gtrsim 1$ and for wave numbers fulfilling $\gamma^2 = k^2\Delta D/D_\text{rot} \gg 1$ in the highly entangled regime
$D_\text{rot}\to 0$ to~\cite{Leitmann:PRL_117:2016}
\begin{align} \label{EqIntScatHermiteApprox}
F(k,t) = e^{-k^2D_\perp t} \sqrt{\frac{\pi}{\gamma}}\frac{1}{\sqrt{2\sinh(2D_\text{rot}\gamma t)}} .
\end{align}
The algebraic decay is observed for times $D_\text{rot} \gamma t \ll 1$ with $F(k,t) = 1/\sqrt{4k^2\Delta D t/\pi}$
until the terminal relaxation $F(k,t) = \sqrt{(\pi/\gamma)} \exp[-(k^2 D_\perp + D_\text{rot}\gamma) t]$ determines the
time-dependent behavior for times $D_\text{rot} \gamma t \gg 1$ [Fig.~\ref{FigCompIntScatClosedExp}].

With the full solution of $F_\text{rod}(k,t)$~[Eq.~\eqref{EqIntScatRodFullSol}] we can assess the validity of the
approximate solution derived by Doi and Edwards for the highly entangled regime:
\begin{align} \label{EqIntScatRodApprox}
F_\text{rod}(k,t) = e^{-k^2 D_\perp t} \int_{-1}^{1}\diff z\
\frac{j_0(kLz/2)^2 e^{-\gamma z^2 \tau_- }}{\sqrt{2\sinh(2 D_\text{rot}\gamma t) \tau_+}},
\end{align}
with $\tau_\pm = [\cosh(2 D_\text{rot} \gamma t) \pm 1]/\sinh(2 D_\text{rot} \gamma t)$. In essence, 
the propagator $G_k$~[Eq.~\eqref{EqSmoluchowskiZPhi}] is approximated by that of a 
harmonic oscillator and the resulting average for the intermediate scattering function is evaluated for a
needle which does not rotate significantly.
As can be inferred from the comparison with the full solution~[Fig.~\ref{FigCompIntScatClosedExp}] the approximate
solution is accurate for intermediate wave numbers, whereas deviations are present for small and high wave numbers. In
particular, for high wave numbers the terminal relaxation is governed by a different prefactor of
$\sqrt{(\gamma/4\pi)}[\int_{-1}^1\diff z\ j_0(kLz/2)\exp(-\gamma z^2/2)]^2$ not captured by the
approximation~[Eq.~\eqref{EqIntScatRodApprox}].

In the presence of other needles~[Fig.~\ref{FigIntScatRodLiqLor}], the validity of the description of the dynamics in terms
of the phantom needle depends on the density and the wave number under consideration. At density $n^* = 128$, which marks
the onset of the scaling behavior of the transport coefficients, the confining tube is only partially present and deviations
become apparent already for wave numbers $kL \gtrsim 10$. Increasing the density by a factor of eight to $n^* = 1024$,
the phantom needle captures the dynamics at much smaller length scales until the dynamics within the tube is resolved
for the largest wave numbers considered. For the needle Lorentz system at density $n^*=1024$, we also observe the
characteristic algebraic decay $\sim t^{-1/2}$ in a small time window for the smallest wave numbers considered.

\section{Summary and conclusion}

We have investigated the dynamics of solutions of infinitely thin needles for densities deep in the semidilute regime. The
needles perform rotational and translational Brownian motion and are not allowed to cross each other. 

From the time dependence of the rotational and translational diffusion, we have extracted the long-time diffusion
coefficients and the geometry of the confining tube emerging for the high entanglement due to neighboring needles.
The transport coefficients as well as the tilt angle and the tube diameter exhibit the predicted density-dependent
scaling behavior obtained from theory and are observed in our simulation over one order of magnitude in the density.

Due to the strong suppression of the rotational and perpendicular translational diffusion coefficients, the mean-square
displacement of the geometric center on coarse-grained time scales for the strong entanglement is purely determined by the diffusion coefficient for
the motion parallel to the long axis. For the non-Gaussian parameter, we observe the pure sliding motion in the confining tube
as an intermediate plateau over many decades in time. 

An analytic expression for the intermediate scattering function of the entire needle has been derived and evaluated
numerically. In comparison to the intermediate scattering function of the geometric center, the characteristic algebraic
decay corresponding to the sliding within the tube is much less pronounced and only observed in a small time window and for
the smallest wave numbers in the high entanglement.

On coarse-grained time and length scales, the phantom needle with long-time translational and rotational diffusion
coefficients as input parameters captures the dynamics of the needles in solution for all the considered quantities as
anticipated from the tube model of Doi and Edwards. We also performed simulations on needle Lorentz systems where a single tracer
needle performs Brownian motion in a quenched array of other needles. The dynamics in needle Lorentz systems is
identical to the dynamics in needle liquids and the dynamic rearrangement due to motion of the other needles only admits 
a change in the absolute values of the long-time diffusion coefficients.

Here, we have considered the transport properties of monodisperse solutions of needles. It would be interesting to go
beyond this model and consider the diffusion of a tracer needle of length $L_\text{t}$ in a solution (matrix) of needles
of a different length $L_\text{s}$.  For needles which perform translational motion only, the scaling behavior of the
perpendicular translational diffusion coefficient of the tracer has been elaborated
theoretically~\cite{Szamel:JCP_100:1994} and a different behavior of $D_\perp \sim (n^*)^{-1}$ is predicted if the
needles of the matrix are much shorter than the tracer needle, $L_\text{s} \ll L_\text{t}$.  In the opposite case
$L_\text{t} \ll L_\text{s}$, where the matrix evolves much slower than the tracer needle which is reminiscent of a 
needle Lorentz system, one theoretically recovers the scaling behavior $D_\perp \sim (n^*)^{-2}$.
Although both cases are computationally more demanding since the averages are performed for the tracer needle only,
it should in principle be possible to assess these predictions in our simulation.  Moreover, one can
consider the dynamics also in the presence of additional spherical particles which affect the diffusion of the needle
parallel to their long axis~\cite{Yamamoto:ACSML_4:2015}.

In the case of hard rods of finite diameter $b$~\cite{Loewen:PRE_50:1994,Cobb:JCP_123:2005,Tao:JCP_124:2006} 
the semidilute regime extends up to densities of $n \ll 1/bL^2$ where the excluded volume
becomes relevant and influences the dynamics. As long as the diameter of the rod is much smaller than its length $b \ll
L$, the scaling behavior of the rotational diffusion coefficient in the semidilute regime is still governed by the same scaling
law~\cite{Doi:JP_36:1975,Cobb:JCP_123:2005,Tao:JCP_124:2006}. For the additional quantities considered here, one
anticipates that the phantom needle still captures the dynamics of such very elongated hard rod solutions on
coarse-grained time and length scales.

One may also relax the requirement of stiff fibers and consider the dynamics in the case of semiflexible
polymers~\cite{Morse:PRE_63:2001,Ramanathan:PRE_76:2007,Nam:JCP_133:2010,Egorov:PRL_116:2016}. While the tube concept
still provides the key insight into the dynamics of such solutions in the highly entangled
regime~\cite{Odijk:MA_16:1983,Semenov:JCSFT_82:1986}, the tube displays additional tube-width
fluctuations~\cite{Glaser:PRL_105:2010,Wang:PRL_104:2010, Glaser:PRE_84:2011} and the tube renewal becomes more complex
which directly affects the behavior of the transport coefficients. However, tube-width fluctuations are not only
important for semiflexible polymers, but also appear to be a relevant concept for needle solutions as has been shown
just recently~\cite{Sussman:PRL_107:2011,Sussman:PRL_109:2012}.

Our simulations of needle Lorentz systems and needle liquids deep in the semidilute regime heavily rely on the use of a
geometry-adapted neighbor list which significantly reduces the computational time for the collision detection. Such
neighbor lists have been considered before for nonspherical particles~\cite{Donev:JCompPhys:2005} and the obtained
speedup for the simulation should at least in part be transferable to simulations of hard semiflexible polymers.

\appendix
\section{Simulation of the hard-core interaction between needles} \label{Appendix:HardCoreInteraction}

We employ a pseudo-Brownian scheme to describe the hard-core interaction between the needles~\cite{Scala:JCP_126:2007,
Hoefling:JCP_128:2008}.
The scheme builds on an event-based algorithm to propagate the needle within one Brownian time step. It extends a
collision detection algorithm developed earlier for a needle moving in a two-dimensional array of point
obstacles~\cite{Hoefling:Phd:2006}.
We always move only a single needle at a time such that the formulas simplify. For the general case, we
refer to Ref.~\cite{Frenkel:MolPhys_49:1983}. 

During a Brownian time step $\Delta t \in [0,\tau_\text{B}]$, the needle
rotates in a plane perpendicular to the rotational pseudovelocity $\boldsymbol\omega$, and the geometric center $\vec{r}(t +
\Delta t) = \vec{r}(t) + \vec{v}\Delta t$ located in the rotational plane moves with pseudotranslational velocity $\vec{v}$. A collision candidate
needle of the same length $L$ with position $\vec{r}_c$ and orientation $\vec{u}_c$ intersects the rotational plane of the
moving needle for times for which the following inequality holds:
\begin{align} \label{EqIntersectionPlane}
|\boldsymbol\omega\cdot\Delta\vec{r}_c(t+\Delta t)| \leq \frac{L}{2}|\boldsymbol\omega\cdot\vec{u}_c|, 
\end{align}
where we defined the distance $\Delta\vec{r}_c(t+\Delta t) = \vec{r}(t+\Delta t) - \vec{r}_c$ of the geometric needle centers.

In addition to that, the intersection point $\vec{r}_I(t+\Delta t)$ of the collision candidate needle with the rotational
plane has to traverse the rotational disk of radius $L/2$ of the moving needle. Defining the distance vector ${\Delta
\vec{r}_\text{pl}(t + \Delta t)} = \vec{r}(t+\Delta t) - \vec{r}_I(t+\Delta t)$ of the geometric center of the moving needle with
the intersection point in the rotational plane, the relevant time interval is determined by the inequality 
\begin{align} \label{EqIntersectionDisk}
[\Delta \vec{r}_\text{pl}(t + \Delta t)]^2 \leq \Bigl(\frac{L}{2}\Bigr)^2 .
\end{align}

Both of the preceding conditions [Eqs.~\eqref{EqIntersectionPlane}~and~\eqref{EqIntersectionDisk}] define a smaller
time interval $[\tau_l,\tau_u] \subset [0,\tau_\text{B}]$, which has to be nonempty for a possible collision of both
needles. A necessary condition for a collision at times $\Delta t\in [\tau_l,\tau_u]$ is that
the distance vector $\Delta \vec{r}_\text{pl}(t + \Delta t)$ and the orientation $\vec{u}(t + \Delta t)$ of the moving needle 
become parallel:
\begin{align}
\boldsymbol\omega\cdot[\Delta \vec{r}_\text{pl}(t + \Delta t) \times \vec{u}(t+\Delta t)] = 0 .
\end{align} 
The scalar product with the pseudorotational velocity $\boldsymbol\omega$ is used both to obtain a one-dimensional equation and
to enforce a change of sign at the root. In order to determine the collision time and also if a collision even occurs,
we use an interval Newton method~\cite{Hansen:Dekker:2004, Hoefling:Phd:2006, Hoefling:PRL_101:2008} on the time interval
$[\tau_l,\tau_u]$. In order not to miss collisions, a correctly rounded math library~\cite{CRlibm_2} has to be used for the
transcendental functions in the change of orientation, $\vec{u}(t+\Delta t)$~[Eq.~\eqref{EqPropagationRules}].
This procedure yields the point in time $t+\tau_\text{c}$ of the next collision.

Upon colliding at time $t + \tau_\text{c}$, the moving needle acquires new velocities for rotation and translation which
are determined by conservation of energy, momentum, and angular momentum. Here, we assume that the center of mass of the
moving needle coincides with its geometric center and consider only smooth
needles~\cite{Frenkel:MolPhys_49:1983,Otto:JCP_124:2006}, where the momentum transfer $\Delta \vec{p} = \Delta p
\vec{e}_m$ is perpendicular to the orientation of both collision partners and directed along $\vec{e}_m =
\vec{u}(t+\tau_c) \times \vec{u}_c/|\vec{u}(t+\tau_c) \times \vec{u}_c|$. The magnitude of the momentum transfer,
$\Delta p$, depends on the translational and rotational velocities $\vec{v}$ and $\boldsymbol\omega$, respectively,
and on the point of contact $\vec{r}_\text{coll} = \Delta \vec{r}_\text{pl}(t + \tau_\text{c})$ with
respect to the center of the moving needle:
\begin{align} 
\Delta p = -2\frac{\vec{v}\cdot\vec{e}_m +
\boldsymbol\omega\cdot(\vec{r}_\text{coll}\times\vec{e}_m)}{1+\frac{m}{I}(\vec{r}_\text{coll}\times\vec{e}_m)^2} .
\end{align}
Then, the new velocities $\vec{v}'$ and $\boldsymbol\omega'$ for translation and
rotation for the remaining Brownian time interval $[\tau_\text{c},\tau_\text{B}]$ are determined by
\begin{align}
\begin{split}
\vec{v}' &= \vec{v} + \Delta p \vec{e}_m,  \\
\boldsymbol\omega' &= \boldsymbol\omega + \frac{m}{I}\Delta p (\vec{r}_\text{coll} \times \vec{e}_m) .
\end{split}
\end{align}

The ratio of mass $m$ and inertia $I$ can be related to the short-time diffusion coefficients by prohibiting an average
flow of energy between the rotational and translational degrees of freedom of the moving needle. We define
pseudotemperatures $T_\text{rot}$ and $T_\perp$ via the relations $I\langle\boldsymbol\omega^2\rangle/2 = 2 k_\text{B}
T_\text{rot}/2$ and $m \langle\vec{v}^2_\perp\rangle = 2 k_\text{B} T_\perp/2$ and determine the remaining averages as
$\langle\boldsymbol\omega^2\rangle = 4 D_\text{rot}^0/\tau_\text{B}$ and $\langle\vec{r}_\perp^2\rangle =
4D_\perp^0/\tau_\text{B}$ from the equations for the pseudovelocities [Eqs.~\eqref{EqPseudoVelocities}]. 
Then, for $T_\text{rot} = T_\perp$, the average flow of energy at collision vanishes and we obtain 
\begin{align}
\frac{m}{I} = \frac{D_\text{rot}^0}{D_\perp^0}.
\end{align}

For anisotropic particles such as needles, it is possible that the moving needle collides again with the previous collision
partner in the remaining time interval $[\tau_\text{c},\tau_\text{B}]$. A minimal propagation time for such an event can
be estimated by considering the moving needle and the intersection point of the collision partner in the new rotational
plane determined by $\boldsymbol\omega'$. Ignoring the length of both needles, the minimal collision time $\tau_\text{r}$ 
for a repeated collision is obtained as a solution of the transcendental equation
\begin{align}
\frac{\omega' \tau_\text{r}}{\tan(\omega' \tau_\text{r})} =  
\omega'\frac{[\Delta \vec{r}_\text{pl}(t + \tau_\text{c}) + \vec{v}_\text{pl}'\tau_\text{r}]
\cdot\vec{u}(t+\tau_\text{c})}{\vec{v}_\text{pl}'\cdot[\boldsymbol\omega'/\omega'\times\vec{u}(t+\tau_\text{c})]} ,
\end{align}
with the velocity of the intersection point in the rotational plane $-\vec{v}_\text{pl}' = -\vec{v}' +
(\boldsymbol\omega'\cdot\vec{v}')\vec{u}_c/(\boldsymbol\omega'\cdot\vec{u}_c)$. By considering the series expansion for
both $\tan(\cdot)$ and $\cot(\cdot)$, the function on the left hand side can be estimated by the inequality $1-(\omega'
\tau_\text{r})^2/2 \leq \omega' \tau_\text{r}/\tan(\omega' \tau_\text{r}) \leq 1-(\omega' \tau_\text{r})^2/3$ valid for
times $0\leq\omega'\tau_\text{r}<\pi/2$. Hence, by solving for two quadratic equations, we obtain a lower bound for the
minimal collision time $\tau_\text{r}$ and only choose $\omega'\tau_\text{r}=\pi/2$ if the solutions are located outside
of the validity of the approximation.

To determine all possible collision candidate needles, we enclose the moving needle in a fixed cylinder, which is valid as long
as the moving needle does not touch the boundary. Then, all needles intersecting the cylinder belong to the class of 
collision candidates. The optimal size of the cylinder has to be determined via the simulation and decreases with
decreasing Brownian time $\tau_\text{B}$. For liquid configurations where we move every needle subsequently, we consider
the intersection of the corresponding cylinders. This geometry-adapted neighbor list significantly reduces the
computational time in the dense regime. 

The collision detection is further supplemented by a simple test performed on the time interval $[\tau_l, \tau_u]$
obtained after checking for both length conditions [Eqs.~\eqref{EqIntersectionPlane}~and~\eqref{EqIntersectionDisk}].
During time $[0, \tau_u]$ the needle covers two sectors of angle $\omega \tau_u$ and we approximate the area by 
two lines parallel to $\vec{u}(t)$ in the rotational plane with minimal distance $L\sin(\omega\tau_u)/2$ to the
geometric center of the moving needle. Then, a collision can only happen if the intersection point traverses this
area in the interval $[\tau_l, \tau_u]$.

\section{Moments of the needle displacement} \label{Appendix:Moments}

Since the intermediate scattering function of the geometric center, $F(k,t) = \langle e^{-\img\vec{k}\cdot\Delta
\vec{r}(t)}\rangle$~[Eq.~\eqref{EqIntScatNeedleCenter}], is isotropic in the wave vector $\vec{k}$, we average over all
possible orientations of the wave vector and obtain an expression which manifestly displays the symmetry:
\begin{align}
F(k,t) = \biggl\langle \frac{\sin(k|\Delta\vec{r}(t)|)}{k|\Delta\vec{r}(t)|}\biggr\rangle ,
\end{align}
with wave number $k = |\vec{k}|$. Then, the mean-square displacement $\text{MSD}(t)$ and the mean-quartic
displacement $\text{MQD}(t)$ of the needle center are encoded in the small-wave-number behavior
\begin{align}\label{EqFSmallWaveNumber}
F(k,t) = 1 - \frac{k^2}{3!}\text{MSD}(t) + \frac{k^4}{5!}\text{MQD}(t) +
\mathcal{O}(k^6) .
\end{align}
Since the full solution of the intermediate scattering function can be expressed in terms of spheroidal wave functions
via
\begin{align} \label{EqInterScatPSn}
F(k,t) = \sum_{\substack{n = 0\\ n\text{ even}}}^\infty \frac{2n + 1}{4} \bigg[\int_{-1}^1\diff{z}\ \Ps_n^0(z,\gamma^2)\bigg]^2 e^{-\Gamma_n^0 t} 
\end{align}
with decay constants $\Gamma_n^0 = D_\parallel k^2 + D_\text{rot}\lambda_n^0$, we use a time-independent perturbation
theory in the wave number $k$, similar to Ref.~\cite{Kurzthaler:SR_6:2016}. We derive the dependence of the spheroidal eigenvalue $\lambda_n^0(\gamma^2)$
and the spheroidal wave function on the real parameter $\gamma^2$ up to order $\mathcal{O}(\gamma^4)$. For the
mean-square and the mean-quartic displacement only $n=0$ and $n=2$ in the intermediate scattering function
$F(k,t)$~[Eq.~\eqref{EqInterScatPSn}] contribute and the expansion for the spheroidal eigenvalue $\lambda_n^0$ reads~\cite{Meixner:Springer:1954}
\begin{align} 
\lambda_n^0(\gamma^2) - n(n+1) = 
\begin{cases}
-\frac{2}{3}\gamma^2-\frac{2}{135}\gamma^4+\mathcal{O}(\gamma^6), & n=0, \\
-\frac{10}{21}\gamma^2+\frac{94}{9261}\gamma^4+\mathcal{O}(\gamma^6), & n=2.
\end{cases}
\end{align}
Similarly for the integral of the spheroidal wave function we obtain
\begin{align}
\int_{-1}^1\diff z\ \Ps_n^0(z,\gamma^2)=
\begin{cases}
2-\frac{1}{405}\gamma^4 + \mathcal{O}(\gamma^6), & n=0, \\
\frac{2}{45}\gamma^2-\frac{4}{2835}\gamma^4 + \mathcal{O}(\gamma^6), & n=2.
\end{cases}
\end{align}
Then, both moments can be obtained by comparing the resulting expression to the small-wave number
behavior~[Eq.~\eqref{EqFSmallWaveNumber}].

\begin{acknowledgments}
The computational results presented have been achieved (in part) using the HPC infrastructure LEO of the University of
Innsbruck. We acknowledge financial support by the Deutsche Forschungsgemeinschaft (DFG) Contract No. FR1418/5-1.
\end{acknowledgments}


%


\end{document}